\documentclass[twocolumn]{aastex63}
\usepackage{amssymb}
\usepackage{amsmath}
\usepackage{apjfonts} 
\usepackage{natbib}

\newcommand{\be}{\begin{equation}}
\newcommand{\ee}{\end{equation}}

\shorttitle{A SUBLIME 3D Model for Cometary Coma Emission}

\begin{document}

\title{A SUBLIME 3D Model for Cometary Coma Emission:\\ the Hypervolatile-Rich Comet C/2016 R2 (PanSTARRS) }

\correspondingauthor{M. A. Cordiner}
\email{martin.cordiner@nasa.gov}

\author{M. A. Cordiner}
\affiliation{Astrochemistry Laboratory, NASA Goddard Space Flight Center, 8800 Greenbelt Road, Greenbelt, MD 20771, USA.}
\affiliation{Department of Physics, Catholic University of America, Washington, DC 20064, USA.}

\author{I. M. Coulson}
\affiliation{East Asian Observatory, 660 N. A’ohoku Place, Hilo, HI 96720, USA.}

\author{E. Garcia-Berrios}
\affiliation{Astrochemistry Laboratory, NASA Goddard Space Flight Center, 8800 Greenbelt Road, Greenbelt, MD 20771, USA.}
\affiliation{Department of Physics, Catholic University of America, Washington, DC 20064, USA.}

\author{C. Qi}
\affiliation{Harvard-Smithsonian Center for Astrophysics, 60 Garden Street, MS 42, Cambridge, MA 02138, USA.}

\author{F. Lique}
\affiliation{Université de Rennes 1, Campus de Beaulieu, 263 avenue du G{\'e}n{\'e}ral Leclerc, 35042 Rennes Cedex, France.}

\author{M. Zo{\l}towski}
\affiliation{Université de Rennes 1, Campus de Beaulieu, 263 avenue du G{\'e}n{\'e}ral Leclerc, 35042 Rennes Cedex, France.}
\affiliation{LOMC - UMR 6294, CNRS-Universit{\'e} du Havre, 25 rue Philippe Lebon, BP1123, F-76 063 Le Havre cedex, France.}

\author{M. de Val-Borro}
\affiliation{Astrochemistry Laboratory, NASA Goddard Space Flight Center, 8800 Greenbelt Road, Greenbelt, MD 20771, USA.}
\affiliation{Department of Physics, Catholic University of America, Washington, DC 20064, USA.}

\author{Y.-J. Kuan}
\affiliation{National Taiwan Normal University, Ting-Chou Road, Taipei 11677, Taiwan, ROC.}
\affiliation{Institute of Astronomy and Astrophysics, Academia Sinica, Taipei 106, Taiwan, ROC.}

\author{W.-H. Ip}
\affiliation{Graduate Institute of Space Science, National Central University, Taoyuan City, Taiwan.}

\author{S. Mairs}
\affiliation{East Asian Observatory, 660 N. A’ohoku Place, Hilo, HI 96720, USA.}

\author{N. X. Roth}
\affiliation{Astrochemistry Laboratory, NASA Goddard Space Flight Center, 8800 Greenbelt Road, Greenbelt, MD 20771, USA.}
\affiliation{Department of Physics, Catholic University of America, Washington, DC 20064, USA.}

\author{S. B. Charnley}
\affiliation{Astrochemistry Laboratory, NASA Goddard Space Flight Center, 8800 Greenbelt Road, Greenbelt, MD 20771, USA.}

\author{S. N. Milam}
\affiliation{Astrochemistry Laboratory, NASA Goddard Space Flight Center, 8800 Greenbelt Road, Greenbelt, MD 20771, USA.}

\author{W.-L Tseng}
\affiliation{National Taiwan Normal University, Ting-Chou Road, Taipei 11677, Taiwan, ROC.}

\author{Y.-L Chuang}
\affiliation{National Taiwan Normal University, Ting-Chou Road, Taipei 11677, Taiwan, ROC.}

\begin{abstract}
The coma of comet C/2016 R2 (PanSTARRS) is one of the most chemically peculiar ever observed, in particular due to its extremely high CO/H$_2$O { and N$_2^+$/H$_2$O ratios}, and unusual trace volatile abundances.  However, the complex shape of its CO emission lines, as well as uncertainties in the coma structure and excitation, has lead to ambiguities in the total CO production rate. We performed high resolution, spatially, spectrally and temporally resolved CO observations using the James Clerk Maxwell Telescope (JCMT) and Submillimeter Array (SMA) to elucidate the outgassing behaviour of C/2016 R2. Results are analyzed using a new, time-dependent, three dimensional radiative transfer code (SUBLIME), incorporating for the first time, accurate state-to-state collisional rate coefficients for the CO--CO system. The total CO production rate was found to be in the range $(3.8-7.6)\times10^{28}$~s$^{-1}$ between 2018-01-13 and 2018-02-01, with a mean value of $(5.3\pm0.6)\times10^{28}$~s$^{-1}$ at $r_H=2.8$--2.9~au. The emission is concentrated in a near-sunward jet, with an outflow velocity $0.51\pm0.01$~km\,s$^{-1}$, compared to $0.25\pm0.01$~km\,s$^{-1}$ in the ambient (and night-side) coma. Evidence was also found for an extended source of CO emission, possibly due to icy grain sublimation around $1.2\times10^5$~km from the nucleus. Based on the coma molecular abundances, we propose that the nucleus ices of C/2016 R2 can be divided into a rapidly sublimating apolar phase, rich in CO, CO$_2$, N$_2$ and CH$_3$OH, and a predominantly frozen (or less abundant), polar phase containing more H$_2$O, CH$_4$, H$_2$CO and HCN.

\end{abstract}


\section{Introduction}

Comets are composed of ice, dust and debris accreted during the epoch of planet formation. Having spent most of their lives frozen, at large distances from the Sun, cometary nuclei contain some of our Solar System's most pristine (thermally unprocessed) material. From studies of their gaseous atmospheres (comae), the properties of the nucleus can be inferred, thereby providing unique insights into the conditions prevalent at the dawn of the Solar System. 

The long-period comet C/2016 R2 (PanSTARRS) is one of the most chemically peculiar comets ever observed. Early observations at a heliocentric distance of around 3 au revealed a visually blue coma with a highly unusual optical spectrum, dominated by CO$^+$ emission, in addition to a rare detection of N$_2^+$ \citep{coc18}, consistent with very strong outgassing of CO and an above average N$_2$/CO ratio in the nucleus. Both CO and N$_2$ sublimate at low temperatures compared to other cometary ices \citep[\emph{e.g.}][]{wom17}, and are therefore considered `hypervolatile'. Their presence in very high abundances { (relative to H$_2$O)} implies that C/2016 R2 formed very cold, and was maintained at temperatures $\lesssim20$~K for the duration of its lifetime. 

Followup observations at infrared and millimeter wavelengths confirmed the presence of an exceptionally CO-rich coma, with a low production rate of H$_2$O (usually the dominant cometary volatile) and extremely unusual abundance ratios for other species compared with the typical cometary population \citep{wie18,biv18,mck19}. Due to their very different sublimation temperatures, the CO to H$_2$O ratio in cometary comae is known to vary as a function of temperature of the nucleus, so some enhancement in CO/H$_2$O is expected at the relatively large heliocentric distance ($r_H$) at which this comet was observed. However, the coma composition of C/2016 R2 clearly differs from other comets observed at similar $r_H$.  Based on Table 7 of \citet{biv18}, the abundance ratios CO:H$_2$O:CH$_3$OH:HCN (normalized to CH$_3$OH) were 94:0.3:1:0.004 in C/2016 R2 at $r_H=2.8$~au, 14:43:1:0.08 in C/1995 O1 at 2.8~au, and 24:27:1:0.10 in C/2006 W3 at 3.2-3.3~au (the H$_2$O abundances in R2 and W3 are from \citealt{mck19} and \citealt{boc10}, respectively), highlighting the CO-richness of the coma, as well as its strong H$_2$O and HCN depletions. Cometary and interstellar ice observations have identified the presence of separate `polar' and `apolar' ice phases, dominated by H$_2$O and [CO + CO$_2$], respectively \citep{mum11b,boo15,lus15}. Studies of coma abundances in C/2016 R2 therefore provide a rare opportunity to investigate volatiles outgassed primarily from the apolar (CO + CO$_2$ dominated) phase, which may provide unique insights into the origin, storage and outgassing mechanisms of the less abundant ices in cometary nuclei.

Complexity of the C/2016 R2 blueshifted CO rotational line profile, as well as uncertainties in the CO excitation calculation, leads to ambiguity in its interpretation, and estimates for the CO production rate (in 2018 January) range from $Q({\rm CO})=(4.6\pm0.4)\times10^{26}$~s$^{-1}$ \citep{wie18} to $(10.6\pm0.5)\times10^{26}$~s$^{-1}$ \citep{biv18}. Limited CO mapping by both studies found marginal evidence for extended CO emission beyond that expected based purely on nucleus-driven outgassing, but this possibility has not yet been investigated in detail. Uncertainties therefore remain regarding (1) the intrinsic CO production rate of this comet, (2) the processes by which CO is released into the gas phase, and (3) the detailed coma morphology.

During 2018 January and February, we undertook a program of time-resolved, high resolution spectroscopy, spatial-spectral mapping and interferometry using the James Clerk Maxwell Telescope (JCMT) and Submillimeter Array (SMA) to elucidate the outgassing behaviour of C/2016 R2. The CO $J=3-2$ and $J=2-1$ lines were observed as a probe of the coma kinetic temperature, and based on the strength of the $^{12}$CO emission in this comet, we searched for the $^{13}$CO isotopologue as a tracer of any unusual isotopic processing in this comet's natal carbon.  We also sought to confirm the comet's peculiar CO:HCN:CH$_3$OH:H$_2$CO ratios through observations of submillimeter rotational lines from these species, and HCO$^+$ was observed as a probe of ion chemistry in the outer coma.

The resulting (spectral-spatial-temporal) dataset is analyzed using a new, time-dependent, three dimensional  radiative transfer code (SUBLIME), which is an evolution of the { steady-state} model used previously by \citet{pag10}, \citet{bog17}, \citet{val18}, and \citet{rot21b}, and includes excitation \emph{via} collisions with CO and electrons, as well as radiative processes. { In Section \ref{sec:unco}, we briefly discuss the conditions for which a time-dependent solution of the molecular excitation produces more accurate results than the steady-state treatment}.  

The emission line profiles from C/2016 R2 are simulated using a two-component (conical jet + ambient coma) outgassing model, allowing the molecular production rates and outflow velocities to vary as a function of coma position. Due to a lack of known collision cross sections, previous models for cometary CO emission have relied on approximations for the CO collisional excitation rates. We therefore performed quantum scattering calculations  (using the coupled-states approximation; \citealt{klo18}) to determine, for the first time, accurate state-to-state collisional (de-)excitation rate coefficients for the CO--CO system. The parameters retrieved from our SUBLIME modeling provides new insights into the coma properties and intrinsic nature of C/2016 R2 (PanSTARRS), one of the most unusual comets of our time.

\section{Observations}

\subsection{James Clerk Maxwell Telescope (JCMT)}

Observations of comet C/2016 R2 (PanSTARRS) were conducted during the period 2018-01-13 to 2018-02-01 using the 15 m JCMT atop Mauna Kea. The comet's position on the sky was tracked using Jet Propulsion Laboratory (JPL) Horizons ephemeris \#14. During this time, the heliocentric distance decreased from $r_H=2.9$ to 2.8~au, the geocentric distance increased from $\Delta=2.1$ to 2.3~au, and the sun-comet-observer (phase) angle increased from $\phi=15^{\circ}$ to $19^{\circ}$.

Spectral line observations were carried out using the RxA3m mm-wave receiver (operating at 212 to 274 GHz) and the HARP 16-element sub-mm focal-plane receiver array (operating at 325 to 375 GHz; \citealt{buc09}).  HARP observations of CO $J=3-2$ were made in two modes: stare and jiggle. During stare observations, the `pointing receptor', H05, tracked the coordinates of the target, while the other receptors (in the square array) each record a spectrum at an offset point in the coma. The $30''$ spacing between receptors means that the target field is under-sampled during stare observations, as a result of the $14''$ telescope beam FWHM at 346~GHz ($20''$ at 230 GHz). Jiggle observations, on the other hand, are designed to obtain spatially complete sampling of the target field. We used the HARP4 jiggle pattern to obtain spectra for each point on a $7''\times7''$ grid, covering a $2'\times2'$ map area. Position switching (offset by $300''$ in azimuth) was performed for the purpose of subtracting spectral contributions from the terrestrial atmosphere and telescope optics. The performance of the individual HARP receptors was monitored throughout the observations, and any sub-optimal receptors were flagged and masked during data reduction.

Most of the spectral data were obtained at a resolution of 31 kHz, across a 250 MHz bandpass, using the ACSIS digital autocorrelation spectrometer. For the $^{13}$CO and C$^{18}$O spectral windows (observed simultaneously), 61 kHz resolution was used, and for CH$_3$OH, a relatively coarse resolution of 448~kHz (over a 1000~MHz bandpass) was used to cover multiple transitions of CH$_3$OH in the $J=7-6$ band at 338~GHz.

Standard observatory amplitude and pointing calibrations were performed at regular (at most, hourly) intervals, demonstrating nominal performance. The all-sky pointing performance of JCMT is $\sim2''$ in each of the (azimuth, elevation) coordinates \citep{cou20}, but telescope tracking accuracy over the course of an hour for a given source is usually better than $1''$. Comparison of the actual telescope tracking position (apparent RA and dec.) with the ephemeris showed tracking errors to be less than $0.1''$ in each coordinate on each night. { The observed antenna temperatures ($T_A^*$) were corrected for sky opacity, forward scattering and spillover. Beam efficiency ($\eta$) was corrected using the standard values of $\eta=0.60$ for RxA3m and 0.64 for HARP. The resulting main beam brightness temperature scale ($T_{MB}$) is expected to be accurate to within $\pm10$\%.}

Table \ref{tab:lines} shows a summary of the JCMT spectral line observations, including the zenith opacity at 225 GHz ($\tau_0$) and on-source integration time per observation (Int.). { Integrated line intensities ($\int{T_{MB}dv}$), and centroid velocities ($\bar{v}$) are also given (with $1\sigma$ uncertainties in parentheses). The uncertainties on $\int{T_{MB}dv}$ include the 10\% intensity calibration error, added in quadrature with the statistical error}.

\begin{table*}
\begin{center}
\caption{Log of JCMT spectral line observations\label{tab:lines}}
\hspace*{-2.5cm}
\begin{tabular}{llr@{$-$}lcccccccc}
\hline
\hline
Date&Species&\multicolumn{2}{c}{Transition}&Obs. Mode&$\tau_0$& Int. &$\int{T_{MB}dv}$ & $\bar{v}$  &$r_H$&$\Delta$&$\phi$\\
&        &       &                              &         &        &  (s) &(mK\,km\,s$^{-1}$)&(km\,s$^{-1}$)&(au)&(au)&($^{\circ}$)\\                   
 \hline
2018-01-13.215&CO&3&2&HARP stare&0.10&600                     &870(92)&$-0.21(0.04)$&2.88&2.13&14.8\\
2018-01-13.253&HCN&4&3&HARP stare&0.10&1200                   &$<78$&---&2.88&2.13&14.9\\
2018-01-13.338&CH$_3$OH&7&6&HARP stare&0.11&900               &73(20)$^a$&$-0.08(0.13)$$^a$&2.88&2.13&14.9\\
2018-01-13.401&HCO$^+$&4&3&HARP stare&0.11&900                &$<93$&---&2.88&2.13&14.9\\
2018-01-13.437&H$_2$CO&$5_{1,5}$&$4_{1,4}$&HARP stare&0.11&900  &$<110$&---&2.88&2.13&14.9\\
2018-01-14.206&CO&2&1&RxA3m&0.17&600                          &542(71)&$-0.18(0.08)$&2.87&2.14&15.1\\
2018-01-14.248&CO&3&2&HARP stare&0.17&120                     &1078(154)&$-0.24(0.10)$&2.87&2.14&15.1\\
2018-01-14.260&CO&3&2&HARP stare&0.17&600                     &926(103)&$-0.22(0.05)$&2.87&2.14&15.1\\
2018-01-14.273&CO&3&2&HARP stare&0.17&120                     &831(131)&$-0.21(0.06)$&2.87&2.14&15.2\\
2018-01-14.333&$^{13}$CO&3&2&HARP stare&0.16&900              &$<229$&---&2.87&2.14&15.2\\
2018-01-14.333&C$^{18}$O&3&2&HARP stare&0.16&900              &$<282$&---&2.87&2.14&15.2\\
2018-01-14.367&CO&3&2&HARP jiggle&0.15&1780                   &834(114)&$-0.25(0.04)$&2.87&2.14&15.2\\
2018-01-14.384&CO&3&2&HARP stare&0.15&120                     &881(140)&$-0.27(0.12)$&2.87&2.14&15.2\\
2018-01-15.197&CO&3&2&HARP stare&0.09&120                     &1084(130)&$-0.21(0.06)$&2.87&2.14&15.4\\
2018-01-15.213&CO&3&2&HARP jiggle&0.08&1780                   &1145(131)&$-0.23(0.02)$&2.87&2.14&15.4\\
2018-01-15.231&CO&3&2&HARP stare&0.08&120                     &1028(117)&$-0.23(0.05)$&2.87&2.14&15.4\\
2018-01-15.253&$^{13}$CO&3&2&HARP stare&0.08&900              &$<84$&---&2.87&2.14&15.4\\
2018-01-15.253&C$^{18}$O&3&2&HARP stare&0.08&900              &$<112$&---&2.87&2.14&15.4\\
2018-01-15.271&CO&3&2&HARP stare&0.08&120                     &1015(114)&$-0.19(0.05)$&2.87&2.14&15.4\\
2018-01-20.247&CO&3&2&HARP stare&0.06&600                     &745(76)&$-0.23(0.03)$&2.85&2.18&16.8\\
2018-01-25.322&CO&3&2&HARP stare&0.14&600                     &881(96)&$-0.23(0.04)$&2.83&2.23&17.9\\
2018-01-31.338&CO&3&2&HARP stare&0.08&600                     &910(93)&$-0.23(0.03)$&2.80&2.29&19.1\\
2018-02-01.278&CO&3&2&HARP stare&0.14&600                     &768(84)&$-0.18(0.04)$&2.80&2.30&19.3\\
\hline
\end{tabular}
\parbox{16cm}{$^a$ Integrated over the three strongest CH$_3$OH transitions (see Section \ref{sec:ul}.)}
\end{center}
\end{table*}

\subsection{Submillimeter Array (SMA)}

Interferometric observations of C/2016 R2 were made on 2018-02-21 using the SMA, when the comet was at $r_H=2.7$~au; $\Delta=2.5$~au; $\phi=21^{\circ}$. Five of the SMA (6~m) antennas were online, and in the subcompact (SUB) configuration at the time of observations, resulting in a spatial resolution of $\approx4.4''\times7.7''$ at 230 GHz. The two SMA receivers were both tuned to 1.3 mm to cover the CO $J=2-1$ transition at 230.538 GHz. The SWARM correlator provides 8~GHz bandwidth per sideband, divided into four equal sized chunks with uniform spectral resolution of 140 kHz (0.18 km\,s$^{-1}$ at 230 GHz). Position and Doppler-tracking of the comet over a 5 hour observing period was performed using JPL Horizons orbital solution \#14, and compensated for by the correlator in real time.

Calibration of visibility phases and amplitudes was performed using periodic observations of quasars 0336+323 and 3c111, at 15 minute intervals.
Measurements of Uranus were used to obtain an absolute scale for calibration of the flux densities. All data were phase- and amplitude-calibrated using
the MIR software package\footnote{http://www.cfa.harvard.edu/$\sim$cqi/mircook.html}. The calibrated SMA data were then exported in uvfits format for subsequent imaging and analysis using the CASA software package version 5.1 \citep{jae08}.

Interferometric imaging and deconvolution was performed using the Hogbom {\tt clean} algorithm, with a $20''$ mask centered on the peak of CO emission and a threshold of twice the RMS noise per channel ($\sigma=0.19$~Jy). Natural visibility weighting was used and the pixel size was set to $0.5''$.

\section{Results}

\subsection{JCMT CO Spectral Line Time Series}

\begin{figure}
\centering
\includegraphics[width=\columnwidth]{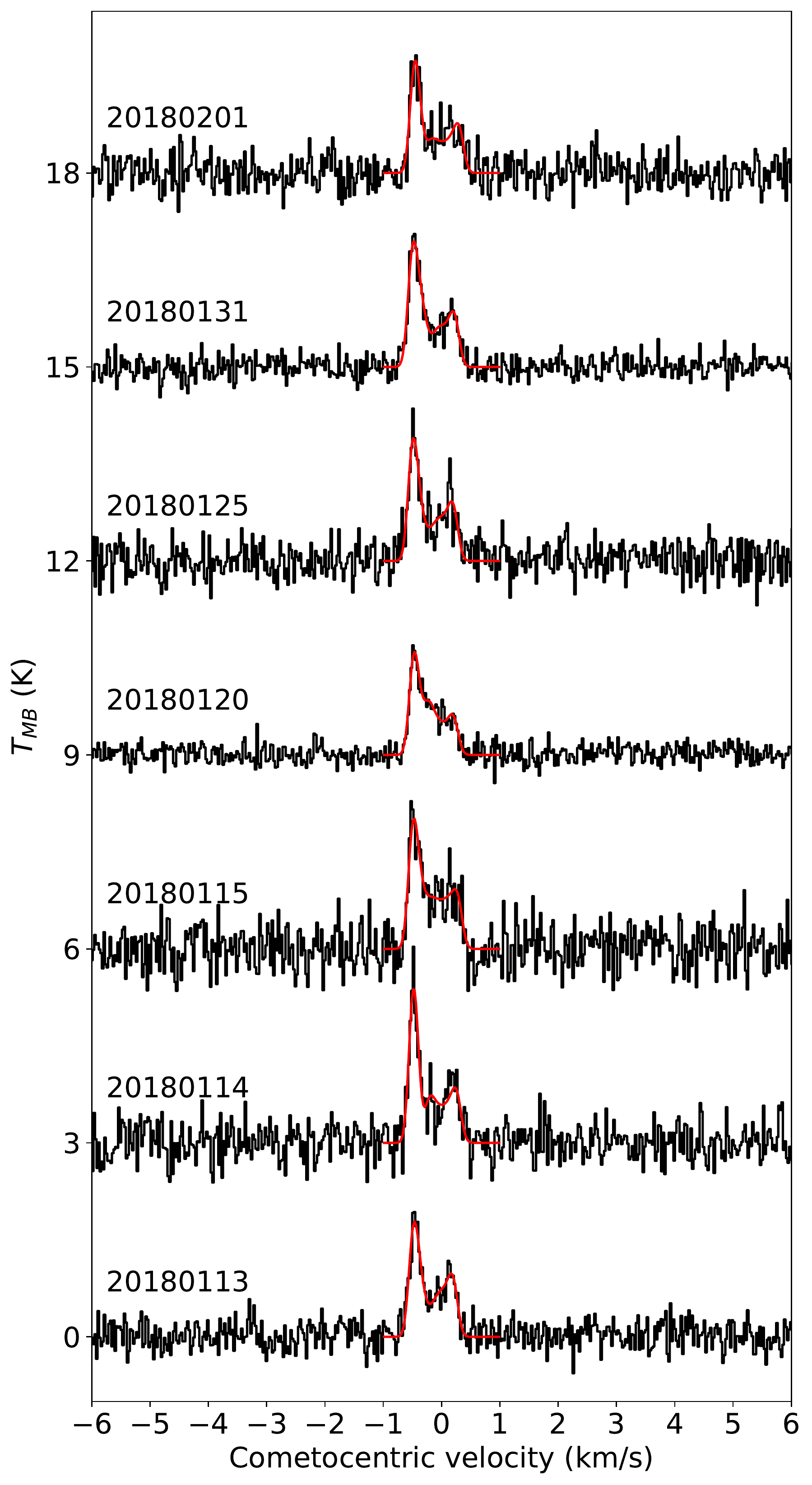} 
\caption{Time sequence of CO $J=3-2$ spectra observed using JCMT (in the topocentric rest frame), shown with additive baseline offsets. The observing date for each spectrum is given in the format YYYYMMDD. For clarity, only spectra with an on-source observing time of at least 600~s are shown. { Best fitting spectral models are overlaid with red curves (see Section \ref{sec:avg}).} \label{fig:co_t}}
\end{figure}

A time series of the observed JCMT CO $J=3-2$ spectra is shown in Figure \ref{fig:co_t}. { Individual spectra have been Doppler-corrected for the comet's radial velocity with respect to the observer, and were baseline-subtracted using low-order polynomial fits to the emission-free regions.} Spectra in this figure were selected based on on-source integration times of at least 600~s, to exclude the noisier data obtained from shorter-integration spectra on some of the observing dates.

The comet's radial velocity in the topocentric rest frame increased steadily throughout the period of observations, but the double-peaked line profile, consisting of a strong, narrow blueshifted peak and a weaker redshifted peak remained apparently constant (within the noise). Given the relatively small phase angle ($\phi$) of our observations, such blueshifted emission is explained as a result of enhanced outgassing from the side of the nucleus facing the Sun (and Earth), compared with the night side.

\begin{figure}
\centering
\includegraphics[width=\columnwidth]{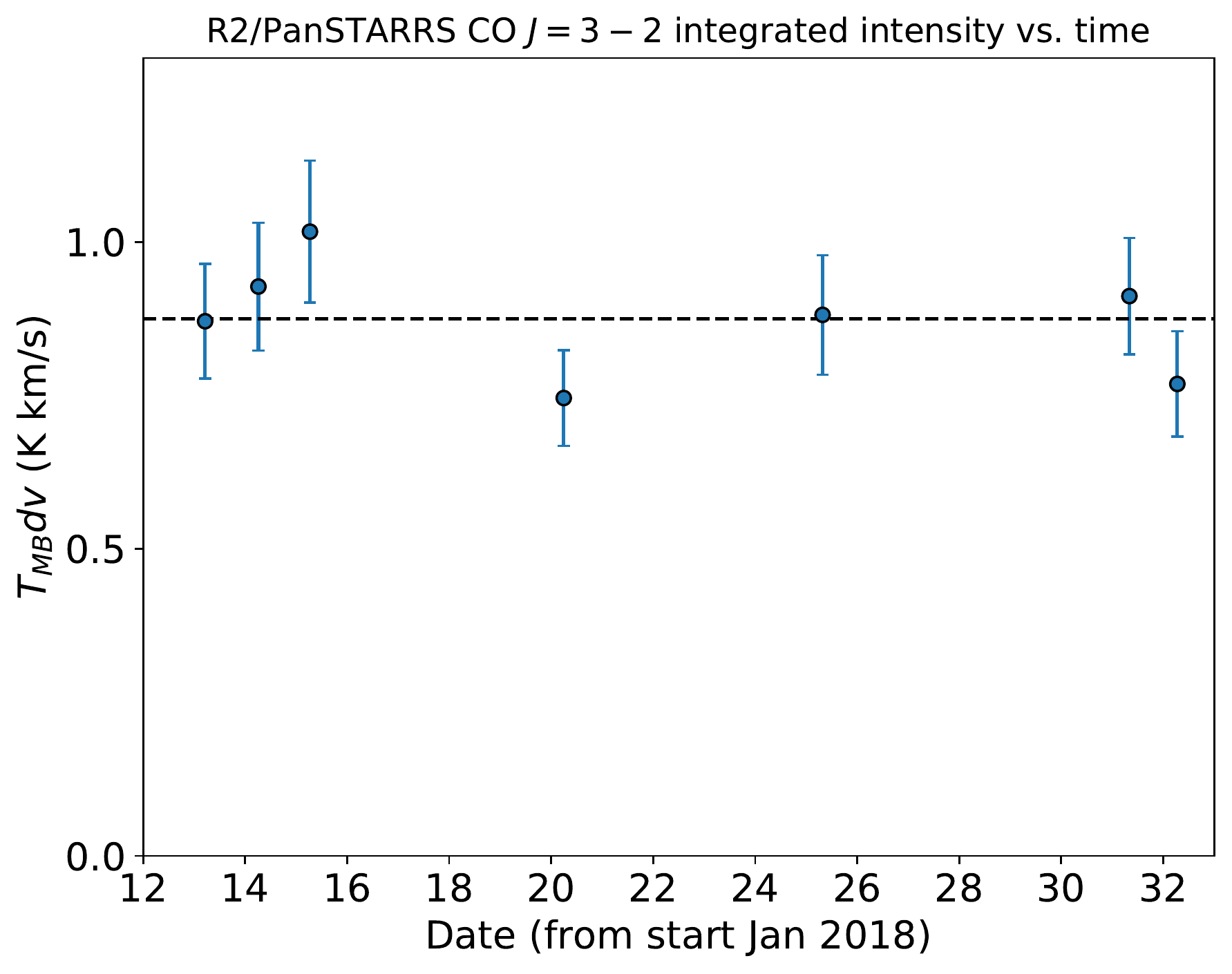} 
\caption{Spectrally-integrated CO $J=3-2$ line intensities observed using the JCMT as a function of time, based on the spectra shown in Figure \ref{fig:co_t}. Horizontal line shows the error-weighted mean. Error bars include the statistical uncertainty, with an additional 10\% calibration uncertainty added in quadrature. \label{fig:co_tdv}}
\end{figure}

To search for temporal variability in the comet's activity, the spectrally integrated emission line area ($\int{T_{MB}dv}$) is plotted as a function of time in Figure \ref{fig:co_tdv}. No significant time-variability was detected, implying that the CO outgassing rate { and rotational temperature ($T_{rot}$) were probably close to steady state during the time period of our observations, although variations in both quantities could have cancelled each other to some degree --- the CO $J=3-2$ line intensity varies by a factor of 1.5 between $T_{rot}=15$--25~K (a plausible range for this comet based on the findings of \citet{biv18}; see also Section \ref{jcmt_co_both})}.

\subsection{CO Single-dish Mapping with JCMT HARP}

\begin{figure}
\centering
\includegraphics[width=\columnwidth]{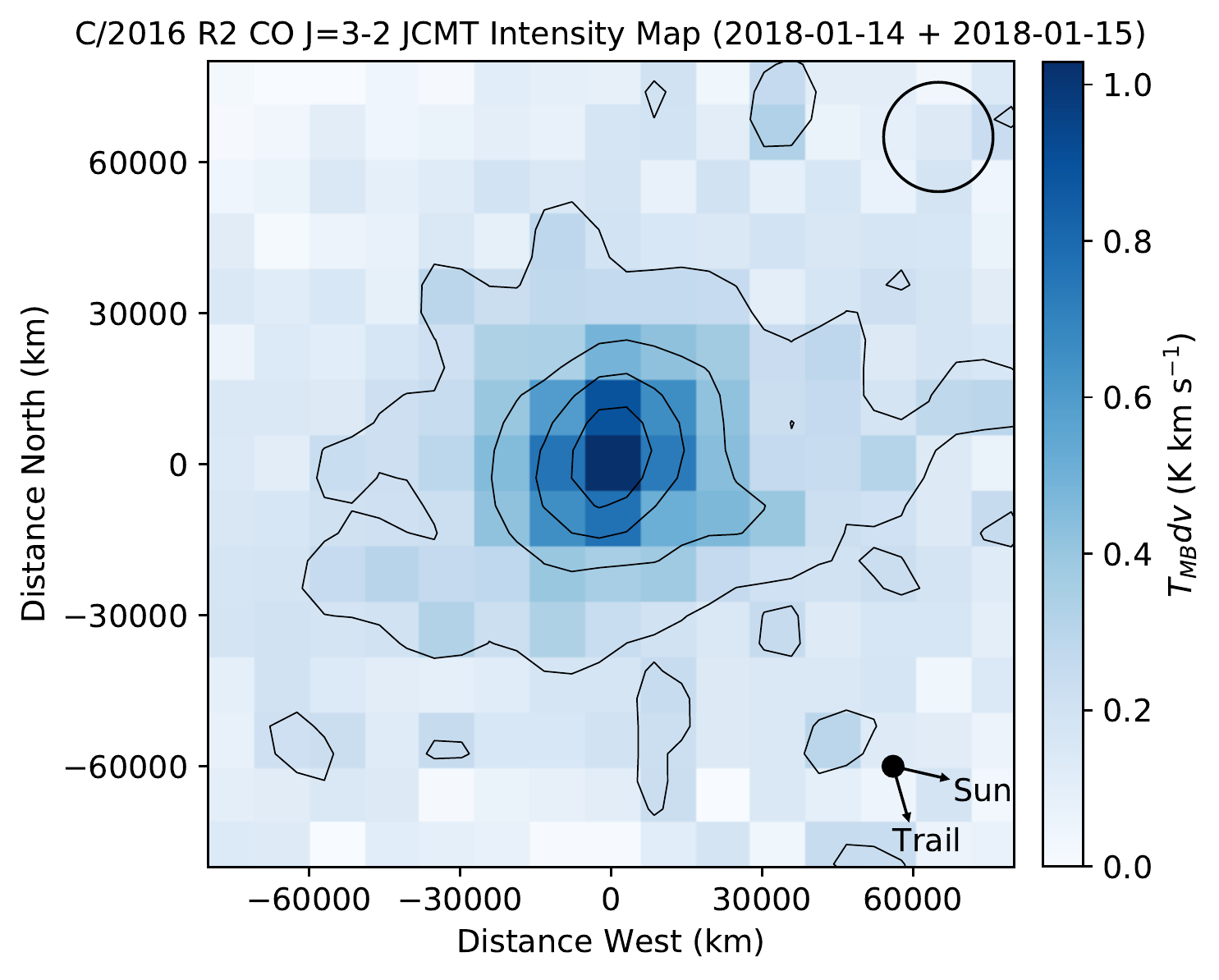} 
\caption{Spectrally integrated CO $J=3-2$ map for comet C/2016 R2 (PanSTARRS), from the average of HARP jiggle observations on 2018-01-14 and 2018-01-15. The FWHM of the circular Gaussian JCMT beam is indicated upper right, and the sky-projected solar and orbital trail (negative of the comet's velocity) vectors are shown lower right. Contours are in units of $3\sigma$ and the axes are aligned with the equatorial (RA/dec.) grid.\label{fig:jcmt_co_map}}
\end{figure}

HARP jiggle-map spectral cubes from 2018-01-14 and 2018-01-15 were Doppler-corrected and averaged together (with rejection of masked pixels). The spectrally integrated (moment 0) map of CO $J=3-2$ emission is shown in Figure \ref{fig:jcmt_co_map}, with contour levels plotted in units of $3\sigma{n_c^{-0.5}}$, where $\sigma$ is the average RMS noise of the data cube (equal to $T_{MB}=72$~mK\,km\,s$^{-1}$) and $n_c$ is the number of spectral channels in the moment 0 integral. The CO $J=3-2$ integrated line intensity, averaged over the entire $120''\times120''$ HARP data cube, { was $164\pm19$ mK\,km\,s$^{-1}$ on 2018-01-14 and $202\pm22$ mK\,km\,s$^{-1}$ on 2018-01-15. These values are sufficiently similar (given the uncertainties) to justify combining the two datasets}. The CO spatial distributions were also apparently identical (within the uncertainties) on the two dates, with consistent radial intensity profiles.

The CO coma is diffuse and spatially extended compared with the 22,000 km JCMT beam, with some { weak} evidence for deviations from circular symmetry about the nucleus. Emission is detected { (at the $3\sigma$ level)} up to a radial distance $82,000$~km west of image-center, which is in a direction similar to the sky-projected sunward vector ($103^{\circ}$ clockwise from North), { whereas the $3\sigma$ contour extends only $58,000$ km to the east}. The 2nd contour (at $6\sigma$) also shows { an excursion} in the sunward direction { into a pixel with an intensity $2.2\sigma$ larger than the mean of the other pixels at the same cometocentric distance}. { These features provide tentative evidence for preferential outgassing on the illuminated (sunward) side of the nucleus}. The (normalized, azimuthally averaged) CO spectral line profile is plotted as a function of distance from the center of the image in Appendix \ref{sec:azavspec} (Figure \ref{fig:azavspec}), and reveals a consistent excess in the blueshifted emission out to cometocentric distances of at least $r_c\sim80,000$~km.

\subsection{CO Interferometric Mapping with the SMA}

\begin{figure}
\centering
\includegraphics[width=\columnwidth]{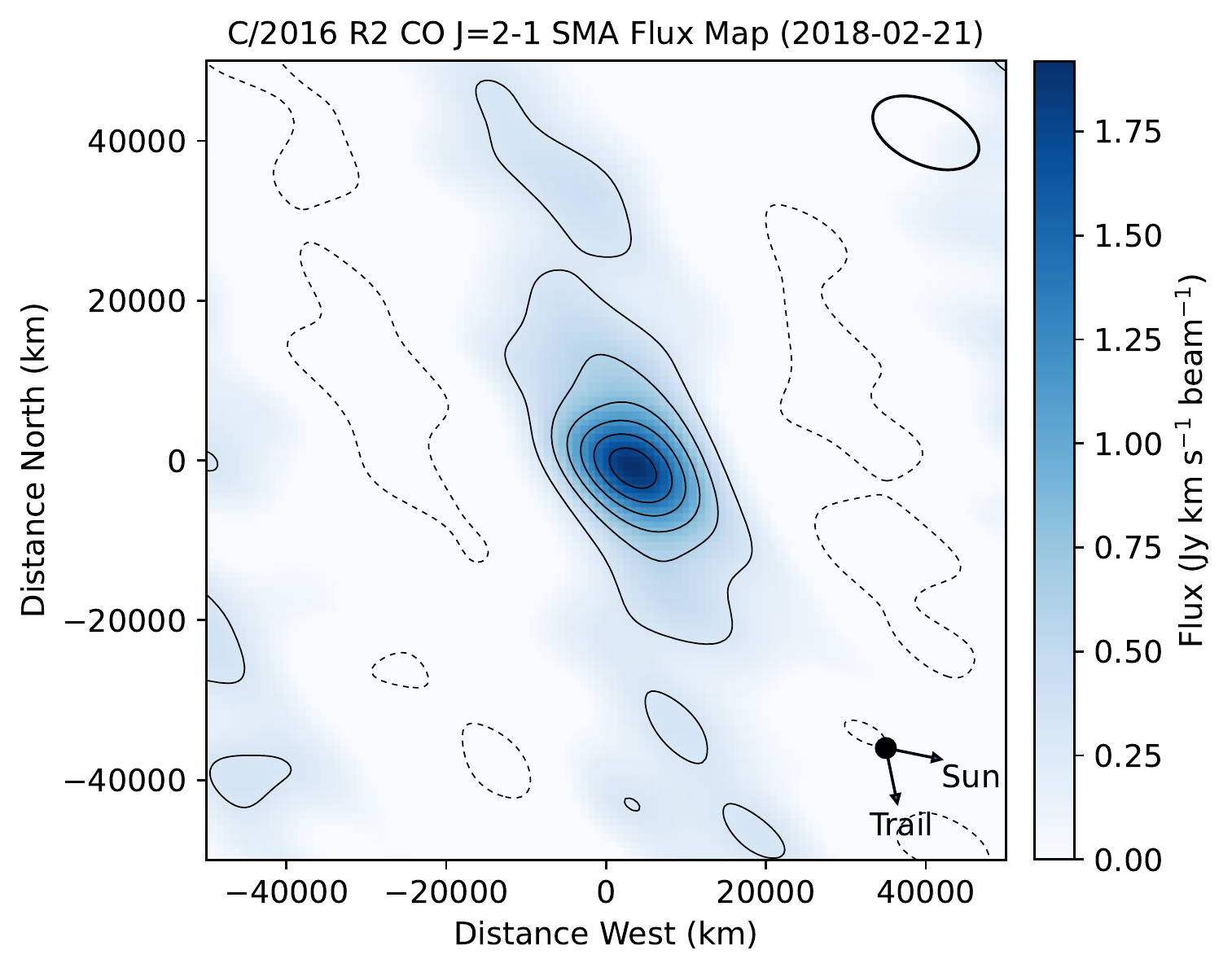} 
\caption{Spectrally integrated CO $J=2-1$ emission map for comet C/2016 R2, obtained using the SMA on 2018-02-21. The FWHM (and orientation) of the elliptical Gaussian restoring beam is indicated by the ellipse (upper right), and the sky-projected solar and orbital trail vectors are shown lower right. Contours are in units of $3\sigma$ and the axes are aligned with the equatorial (RA/dec.) grid, with the origin at the SMA phase tracking center. Negative contours are shown with a dashed line style.\label{fig:sma_co_map}}
\end{figure}

The CO $J=2-1$ intensity map observed using SMA is shown in Figure \ref{fig:sma_co_map}, integrated over the velocity width of the detected emission. The intensity reaches a peak $2.5''$ west of the phase center, which may be { partly a result of asymmetrical outgassing in the sunward direction, or errors} in the position of the comet nucleus compared with the JPL Horizons ephemeris orbital solution.

The coma shows an extended morphology in an approximately N-S direction, and is less well resolved in the E-W direction, where significant large-scale flux appears to have been resolved out by the interferometer, resulting in negative side lobes (regions with dashed contours) apparent on either side of the comet. Although the orientation of the spatially-extended emission (defined by the outermost, $3\sigma$ contour) matches closely the direction of the comet's (sky-projected) orbital trail, it also aligns with an axis of strong artifacts in the interferometric point spread function, so the reality of this asymmetric, extended feature remains questionable.

\section{Radiative Transfer Modeling}

As a result of near-spherical expansion, cometary comae span an extremely broad range of densities over a short distance. Consequently, their gases are often subject to a range of excitation conditions within a single telescope beam, governed by a balance of microscopic collisional and radiative processes \citep{boc04}, and are generally not in local thermodynamic equilibrium (LTE). To interpret cometary rotational spectra therefore requires detailed excitation and radiative transfer modeling. Here we introduce a new code called SUBLIME (SUBlimating gases in LIME) for simulating the rotational emission lines from cometary coma molecules in three dimensions (two spatial and one spectral), where LIME refers to the LIne Modeling Engine by \citep{bri10}.  

The basic equations of radiative transfer and excitation used in our model are described in Appendix \ref{sec:radtran}. Some recently published coma radiative transfer models have invoked the steady-state approximation \citep[\emph{e.g.}][]{bog17,val18,cor19}, setting $dN_i/dt=0$ in Equation \ref{eq:excite}, which allows the energy level populations to be solved independently (in parallel) at a large number of discrete positions within the region of interest. This approximation facilitates the treatment of complex (3D) coma morphologies, but comes at the expense of discarding the effects of the outflow dynamics, which can be important for molecules such as CO, with small dipole moments and hence slow rotational transitions relative to the dynamical timescale (see Section \ref{sec:unco}). Our radiative transfer code is based on the open-source version 1.9.3 of LIME (available at https://github.com/lime-rt/lime/releases). LIME, however, employs the steady state approximation.

We therefore modified substantially the LIME code to enable the more physically accurate, time-dependent solution of Equation \ref{eq:excite}, using the CVODE solver \citep{hin19} to calculate the molecular excitation along radial trajectories of the outflowing coma gases. { The time-dependent solution has previously been implemented in the models of \citet{boc87} and \citet{biv99}, and} allows the temporal evolution of molecular excitation in the rapidly expanding coma to be properly accounted for. { The time-dependent version of our code, as used in the present study, ignores the impact of opacity/photon trapping on the molecular excitation, which is negligible for the molecules considered in this study (see Appendix \ref{sec:radtran}).} 

SUBLIME calculates the molecular excitation on an unstructured 3D grid \citep{del34}, which can be configured with a density of grid points proportional to the gas density. The broad range of density and size scales in the coma (covering many orders of magnitude, from the $\sim$ km-sized nucleus to the $\sim10^6$ km-scale outer coma) can thus be sampled much more efficiently than with a uniform grid spacing. The final level populations $N_i$ as a function of radius are mapped onto the Delaunay grid for raytracing. To simulate coma asymmetries and jets using this method, the spatial domain is divided into multiple solid-angle regions ($\Omega_i$), each with its own outflow velocity ($v_i$), kinetic temperature ($T_i$) and molecular production rate ($Q_i$). A separate CVODE calculation is performed for each solid angle region. In the present article, we use two regions, corresponding to (1) the ambient coma and (2) a conical jet with its apex at the center of the nucleus. 

For all models, we used a Delaunay grid with 10,000 points, subject to a logarithmic density distribution with radius between $r_c=500$~m (the assumed radius of the nucleus) and $10^6$ km. Tests showed that this grid density was sufficient to produce reliable spectral line models at the resolution of our observations; adding more grid points did not significantly change the results.  The precise grid point locations are selected pseudo-randomly for each model run \citep{bri10}, but we fixed the random number generator seed so that an identical grid was produced every time, providing numerical stability in order to facilitate reliable (repeatable) parameter retrievals. During raytracing, we employed the LIME {\tt traceray\_smooth} algorithm, which interpolates the level populations between grid points, thus reducing grid-related artifacts in the output image. A pixel size of $0.5''$ and channel spacing of 25~m\,s$^{-1}$ were chosen for the model images to sufficiently sample the spatial and spectral resolution elements of the JCMT and SMA observations. To accurately capture the rapid (nonlinear) flux increase on sub-pixel scales towards the center of the image, due to the strongly increasing coma density with decreasing $r_c$, we employed cartesian supersampling (on a regularly-spaced grid of $30\times30$ rays) for each pixel within the central $4\times4$ pixel region of each image.

\subsection{CO--CO Collision Rate Coefficients}
\label{sec:rates}

As the dominant coma gas \citep{mck19}, CO is the primary collision partner in our model, and is therefore the main species responsible for the redistribution of thermal energy among the rotational states ($J$) of the observed gases. Knowledge of the CO--CO collision rate coefficients ($k_{J_{1}J_{2}}$) is therefore required to correctly model the CO emission from the comet. Previous studies \citep{wie18,biv18} made gross approximations for these rates, so their results remain uncertain. Here we employ quantum calculations to model the CO--CO collisions, allowing us to accurately determine the CO excitation for the first time in a CO-dominated cometary coma.\footnote{\citet{nde15} previously studied the energy transfer in CO--CO collisions, but their results were only partly converged and contain a systematic error for some (but not all) transition rates.}
 
To describe the interaction between colliding CO molecules, we used the 4D potential energy surface (PES) with rigid CO molecules calculated by \citet{vis03}. The PES was calculated using the coupled cluster single, double and perturbative triples (CCSD(T)) method with augmented triple zeta basis (aug-cc-pVTZ). The accuracy of the PES was benchmarked with respect to experimental studies (\citealt{sur07}, \citealt{sun20}). The scattering calculations were performed with the {\tt Moslcat} code under the assumption of distinguishable particles \citep{mol94}. A series of tests was performed, revealing that the most accurate close-coupling (CC; \citealt{gre75}) approach would not be feasible in terms of computer memory and processing time. We therefore explored the possibility of using the coupled-states (CS) approximation \citep[see][for a review of these methods]{klo18}. The differences between CC and CS were found, on average, to be less than a factor of 1.5--2 (and never higher than a factor of 3).

Assuming that we can distinguish the CO molecules, the first one is the `target' (with rotational state characterized by quantum number $J_1$), and the second is the `collider' ($J_2$). Collisional rate coefficients $k_{J_{1}J_{2} \to J_1'j_2'}$ were then computed by averaging the cross-sections over the Boltzmann distribution of collisional energies (Equation \ref{eq1}). 

\begin{equation}
\label{eq1}
    k_{J_{1}J_{2} \to J_1'J_2'}(T) = \left(\frac{8}{\pi\mu k^{3}_{B}T^{3}}\right)^{\frac{1}{2}} \int\limits_{0}^{\infty} \sigma_{J_{1}J_{2} \to J_1'J_2'}(E_c) E_ce^{-\frac{E_c}{k_{B}T}} dE_c 
\end{equation}

where $\mu$ is the reduced mass of the system, $k_{B}$ is a Boltzmann constant, $\sigma$ is the cross-section of a given transition, and $E_{c}$ is collisional energy.  Rate coefficients used in the radiative transfer model ($k_{J_1 \to J_1'}(T_{kin})$; see Appendix \ref{sec:ratestab}, Table \ref{tab:rates}) were calculated by averaging over a thermal rotational distribution for the initial excitation state of the collider, and summed over its final state as follows:

\begin{equation}
\label{eq2}
\begin{split}
    k_{J_{1} \to J_1'}(T) = N_{J_{2}=0} \sum_{J_2'} k_{J_1,J_2 = 0 \to J_1',J_2'}\\
     + N_{J_{2}=1} \sum_{J_2'} k_{J_1,J_{2} = 1 \to J_1',J_2' } + ...
\end{split}
\end{equation}

where { summation is} over all possible final states of the collider. $N_{J_2}$ is the (Boltzmann) energy level population of the collider, given by

\begin{equation}
\label{eq3}
   N_{J_2} = \frac{(2J_{2}+1)e^{\frac{-E_{J_{2}}}{k_{b}T}}}{\sum_{J_2}(2J_{2}+1)e^{\frac{-E_{J_{2}}}{k_{b}T}}}
\end{equation}

where $E_{J_{2}}$ is its energy level. Calculations were performed up to a collision energy of 300 cm$^{-1}$ and maximum rotational quantum number $J_1 = J_2 = 5$, leading to rate coefficients for kinetic (and rotational) temperatures valid up to 30 K. Such a restricted set of rate coefficients is appropriate for modeling the CO excitation in C/2016 R2 due to the relatively low coma kinetic temperature ($\sim20$~K; \citealt{biv18}); at this temperature, energy levels above $J=5$ comprise less than 1\% of the total CO population in the collisionally-dominated zone. { Our model for C/2016 R2 includes levels up to $J=40$, but for $J>5$, the CO--CO collision rates are assumed to be the same as for para-H$_2$ colliding with CO \citep{yan10}}. A more detailed explanation of the collision rate calculation method, including calculation of rate coefficients for kinetic temperatures up to 100 K, is currently in preparation.

\subsection{Modeling the JCMT CO $J=2-1$ and $J=3-2$ Data}
\label{jcmt_co_both}

On 2018-01-14, the CO $J=2-1$ and $J=3-2$ lines were observed in close succession (within 1.3~hr of each other). Given the lack of significant temporal variability in the $J=3-2$ line strength around this date (Figure \ref{fig:co_t}), differences between the strengths of these two lines can be assumed to originate as a result of (1) their different intrinsic line strengths and (2) differences in the excitation of the upper-state energy level. According to Equation \ref{eq:excite}, the level populations depend on the collision rates $k_{ij}n$ { (where $n$ is the local density)}, which in turn depend on the CO production rate, coma outflow velocity and kinetic temperature.  Modeling the two CO lines simultaneously therefore provides a diagnostic of the coma temperature, while the line profile provides information on the coma outflow velocity along the line of sight. 

There is insufficient information from these relatively low signal-to-noise, 1D spectra to infer the entire coma physical structure, so we adopt a modified \citet{has57} model. The assumption of isotropically expanding gas, with a constant production rate and outflow velocity (as described by the 1D Haser model), is routinely used for analysis of cometary spectra and images observed across the range of wavelengths \citep[\emph{e.g.}][]{boc04,coc12,cor14}. In the case of strongly asymmetric outgassing, however, as observed for C/2016 R2, the assumption of spherical symmetry is no longer applicable, so a more flexible, 3D model is required.

In the present study, we adopt the simplest physically-reasonable model capable of fitting the observed data. We consider two different outflow components: C$_1$ and C$_2$, corresponding to solid angle regions $\Omega_1$, $\Omega_2$ with independent production rates ($Q_1$, $Q_2$) and outflow velocities ($v_1$, $v_2$).  This assumes that the nucleus can be divided into two different activity regimes: (1) an ambient outflow from the majority of the (thermally-activated) sublimating area of the nucleus, and (2) enhanced gas production from a (set of) spatially-confined vent(s) or jet(s), in particular, on the sunward-facing side of the nucleus. Similar 2-component models (involving a sunward jet and ambient/isotropic coma, or a hemispherically asymmetric outflow) have been invoked previously to explain asymmetries in high-resolution spectral line profiles from several comets, including 29P/SW1 \citep{fes01}, O1/Hale-Bopp \citep{gun03}, 19P/Borelly \citep{boc04b}, 2I/Borisov \citep{cor20}, 46P/Wirtanen \citep{rot21} and C/2015 ER61 \citep{rot21b}. The increased production rate and outflow velocity measured on the sunward side of these comets is consistent with the results of fluid dynamic and Monte Carlo coma models \citep{cri99,fou16}, and arises as a result of elevated temperatures in both the coma and the nucleus, due to increased solar insolation.  While more complex coma parameterisations can be envisaged, the simplest model capable of fitting the data should be the best-constrained, with the additional benefit of being more efficient to configure and run.

We construct a SUBLIME model to simultaneously fit the CO $J=2-1$ and $J=3-2$ line profiles, assuming a constant coma kinetic temperature (after \citealt{biv18}). Component C$_1$ is defined by a conical region about the subsolar point, with half-opening angle $\theta$, whereas C$_2$ is the remaining, ambient coma. We used the MPFIT nonlinear least-squares routine \citep{mar12} to find the optimal set of parameters ($Q_1$, $Q_2$, $v_1$, $v_2$, $\theta$). The model images were convolved with the (Gaussian) JCMT beam pattern and normalized by their respective beam efficiency factors before comparing them with observations. Statistical ($1\sigma$) error estimates are obtained for each parameter from the diagonal elements of the MPFIT covariance matrix.

\begin{figure}
\centering
\includegraphics[width=\columnwidth]{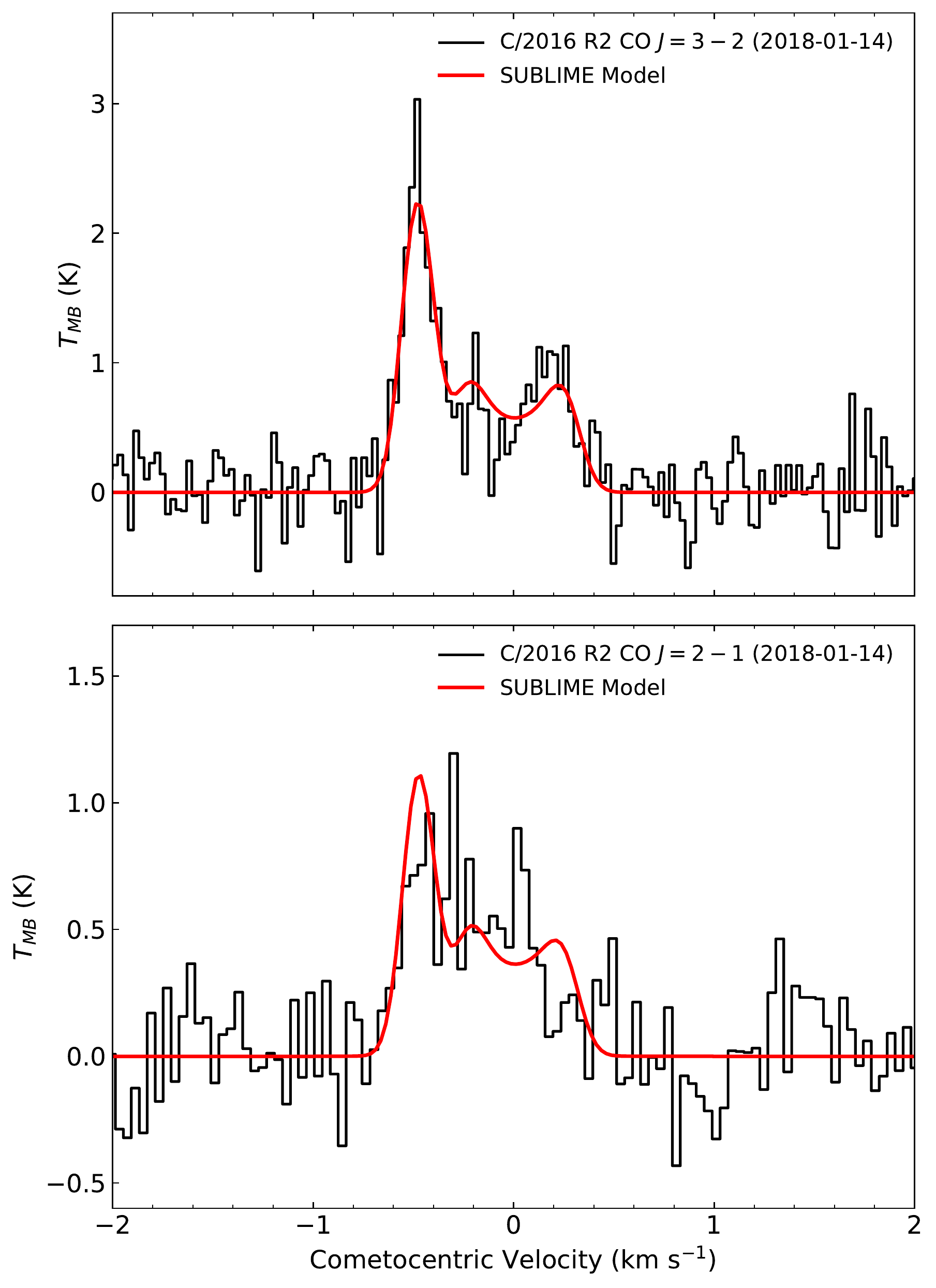} 
\caption{JCMT spectra of CO $J=3-2$ (top) and $J=2-1$ (bottom) (in the cometocentric rest frame). The best-fitting SUBLIME model is overlaid with red curves  \label{fig:co_both}}
\end{figure}

The best fitting model spectra are shown in Figure \ref{fig:co_both}, corresponding to $Q_1=(1.8\pm0.4)\times10^{28}$~s$^{-1}$, $Q_2=(2.2\pm0.3)\times10^{28}$~s$^{-1}$, $v_1=0.50\pm0.01$~km\,s$^{-1}$, $v_2=0.29\pm0.02$~km\,s$^{-1}$, $\theta=27^{\circ}\pm5^{\circ}$, and $T_{kin}=18.7\pm2.3$~K. The retrieved kinetic temperature is consistent with the values $18.6\pm2.6$ to $24.2\pm7.9$ derived by \citet{biv18} using multiple lines of CH$_3$OH on 2018 January 23-24. The total CO gas production rate from our model is $Q_t({\rm CO})=(3.9\pm0.7)\times10^{28}$~s$^{-1}$, and the ratio of production rates per unit solid angle between the jet and ambient coma is $R_Q=(Q_1/\Omega_1)/(Q_2/\Omega_2)=14.5\pm3.9$. Given the strong dependence of the production rate for both components on the size of their solid angle regions, $R_Q$ is a more physically meaningful quantity than the simple ratio of production rates ($Q_1/Q_2$), revealing the degree to which the comet's activity is enhanced due to heating by the Sun in the vicinity of the sub-solar point.  The overall quality of fit is good considering the noise, and reproduces well the asymmetry of the $J=3-2$ line. We also performed fits allowing the angle of the jet axis to vary with respect to the Sun-comet vector, but the quality of fit was not significantly improved. { Additional models were run allowing different $T_{kin}$ values for the two coma components, with best-fitting results $T_{kin_1}=19.3\pm3.7$~K and $T_{kin_2}=17.9\pm4.8$~K, but again, the overall quality of fit was not improved. The consistency of $T_{kin_1}$ and $T_{kin_2}$ (within their respective errors) provides further justification for adopting a uniform $T_{kin}$ value throughout the rest of this study.}

\subsection{Modeling the Average and Time Series JCMT CO $J=3-2$ Spectra}
\label{sec:avg}

\begin{figure}
\centering
\includegraphics[width=\columnwidth]{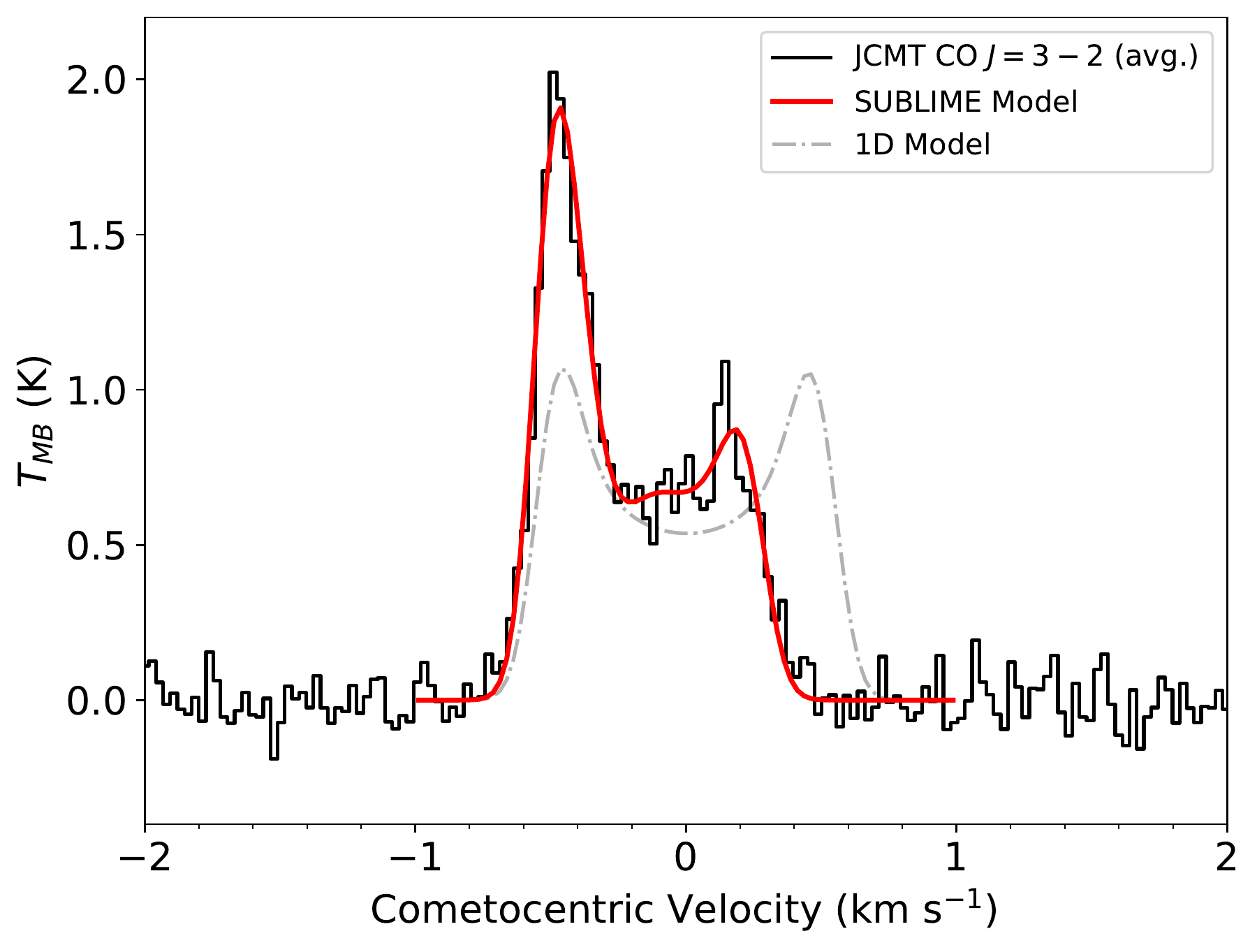} 
\caption{Weighted average CO $J=3-2$ spectrum from the JCMT HARP pointing receptor (in the cometocentric rest frame). The best-fitting SUBLIME 3D model is overlaid using a red curve. Dot-dashed grey curve shows the best-fitting 1D model, assuming a spherically-symmetric coma.  \label{fig:co_avg}}
\end{figure}

Given the relative constancy of the comet's CO emission over time, a useful estimate for the total (time-averaged) CO production rate over the course of our observations was obtained by taking the average of all the JCMT CO ($3-2$) spectral data listed in Table \ref{tab:lines}. After Doppler correcting each spectrum to the cometocentric rest frame and weighting them by $1/\sigma^2$, the resulting average spectrum was modeled using the same procedure described above. The best-fitting model is shown in Figure \ref{fig:co_avg}, and corresponds to $Q_t({\rm CO})=(5.3\pm0.2)\times10^{28}$~s$^{-1}$, $R_Q=7.4\pm0.4$, $v_1=0.51\pm0.01$~km\,s$^{-1}$, $v_2=0.25\pm0.01$~km\,s$^{-1}$, $\theta=62^{\circ}\pm2^{\circ}$. { In this model the direction of the jet's axis with respect to the line of sight ($\phi_{jet}$) was also allowed to vary, and the best fit was for $\phi_{jet}=24^{\circ}\pm2^{\circ}$ ($7^{\circ}\pm2^{\circ}$ away from the mean sun-comet vector)}. These values are within $2\sigma$ of the parameters derived for the simultaneous fit to the $J=2-1$ and $J=3-2$ data on 2018-01-14. Allowing for an additional 10\% calibration error, the total uncertainty on the CO production rate is $0.6\times10^{28}$~s$^{-1}$.

Individual fits were also performed to the time series of spectra presented in Figure \ref{fig:co_t}. A good fit was obtained in each case, and the results are given in Table \ref{tab:fits}. While the majority of these individual fit results are consistent (within 1--2 $\sigma$) with the average JCMT spectrum, notable deviations include the significantly narrower jet opening angle of $\theta=32^{\circ}\pm7^{\circ}$ on January 14th. On February 1st, the ambient coma showed an enhanced outgassing velocity (with $v_2=0.33\pm0.02$~km\,s$^{-1}$), which was also accompanied by a reduction in $R_Q$ to $4.9\pm0.9$. Such variability in the detailed outflow morphology on specific epochs implies some inhomogeneity of the nucleus (and/or its heating rate), leading to modest changes in the CO outgassing behaviour over time.

\begin{table*}
\centering
\caption{Best-fit model results for individual JCMT (and SMA) spectra of C/2016 R2 \label{tab:fits}}
\begin{tabular}{lclccll}
\hline\hline
Date	&	$Q_t$ &	\multicolumn{1}{c}{$R_Q$}	& $v_1$ & $v_2$ & \multicolumn{1}{c}{$\theta$} & \multicolumn{1}{c}{$\phi_{jet}$}\\
      & ($10^{28}$\,s$^{-1}$) & & (km\,s$^{-1}$) & (km\,s$^{-1}$) & \multicolumn{1}{c}{(deg.)} & \multicolumn{1}{c}{(deg.)}\\
\hline
2018-01-13	&4.5 (0.7)	&6.8	(4.5) &0.51	(0.01)	&0.23	(0.02)	&60	(5)  &18	(6)\\
2018-01-14	&4.0 (0.9)	&10.8	(2.9) &0.50	(0.01)	&0.29	(0.02)	&32	(7)  &7	(10)	\\
2018-01-15	&7.4 (1.6)	&9.6	(3.1) &0.52	(0.02)	&0.31	(0.03)	&77	(13) &37	(19)	\\
2018-01-20	&3.9 (1.8)	&17.2	(10.8)&0.50	(0.01)	&0.24	(0.01)	&42	(19) &34	(11)	\\
2018-01-25	&5.0 (0.9)	&10.4	(1.7) &0.53	(0.01)	&0.24	(0.02)	&58	(6)  &20	(7)\\
2018-01-31	&5.7 (0.8)	&7.3	(0.7) &0.53	(0.01)	&0.26	(0.01)	&64	(3)  &12	(4)\\
2018-02-01	&4.4 (0.8)	&4.9	(0.9) &0.49	(0.01)	&0.33	(0.02)	&46	(8)  &8	(12)	\\[2mm]
JCMT avg.	&5.3 (0.6)	&7.4	(0.4) &0.51	(0.01)	&0.25	(0.01)	&62	(2)  &24	(2)\\[2mm]
2018-02-21$^a$ &6.7 (0.9)&	9.2	(1.5)& 0.64	(0.02)&	0.20	(0.02)&	74	(7)&  21$^b$  \\
\hline
\end{tabular}
\parbox{\textwidth}{Notes --- $1\sigma$ statistical uncertainties on each value are given in parentheses. { Uncertainties on $Q_t$ include a 10\% intensity calibration error, added in quadrature with the statistical error.}  $^a$SMA best-fitting parent visibility model. $^b$Jet axis set to direction of the comet-sun vector due to a lack of constraints caused by the lower spectral resolution of the SMA data.}
\end{table*}

\subsection{Modeling the CO Spatial Distribution}
\label{sec:harp}

\begin{figure*}
\centering
\includegraphics[width=0.32\textwidth]{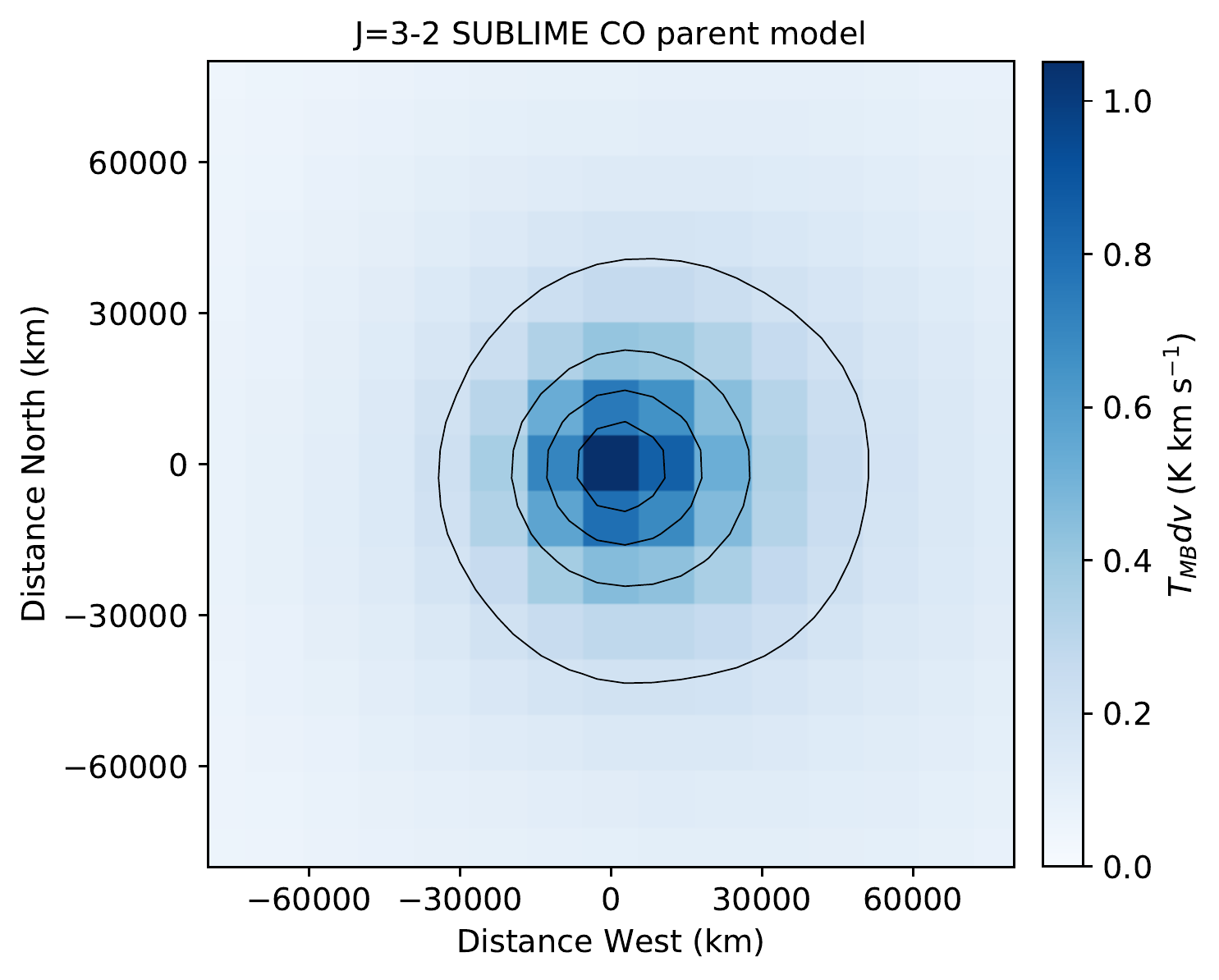}
\includegraphics[width=0.32\textwidth]{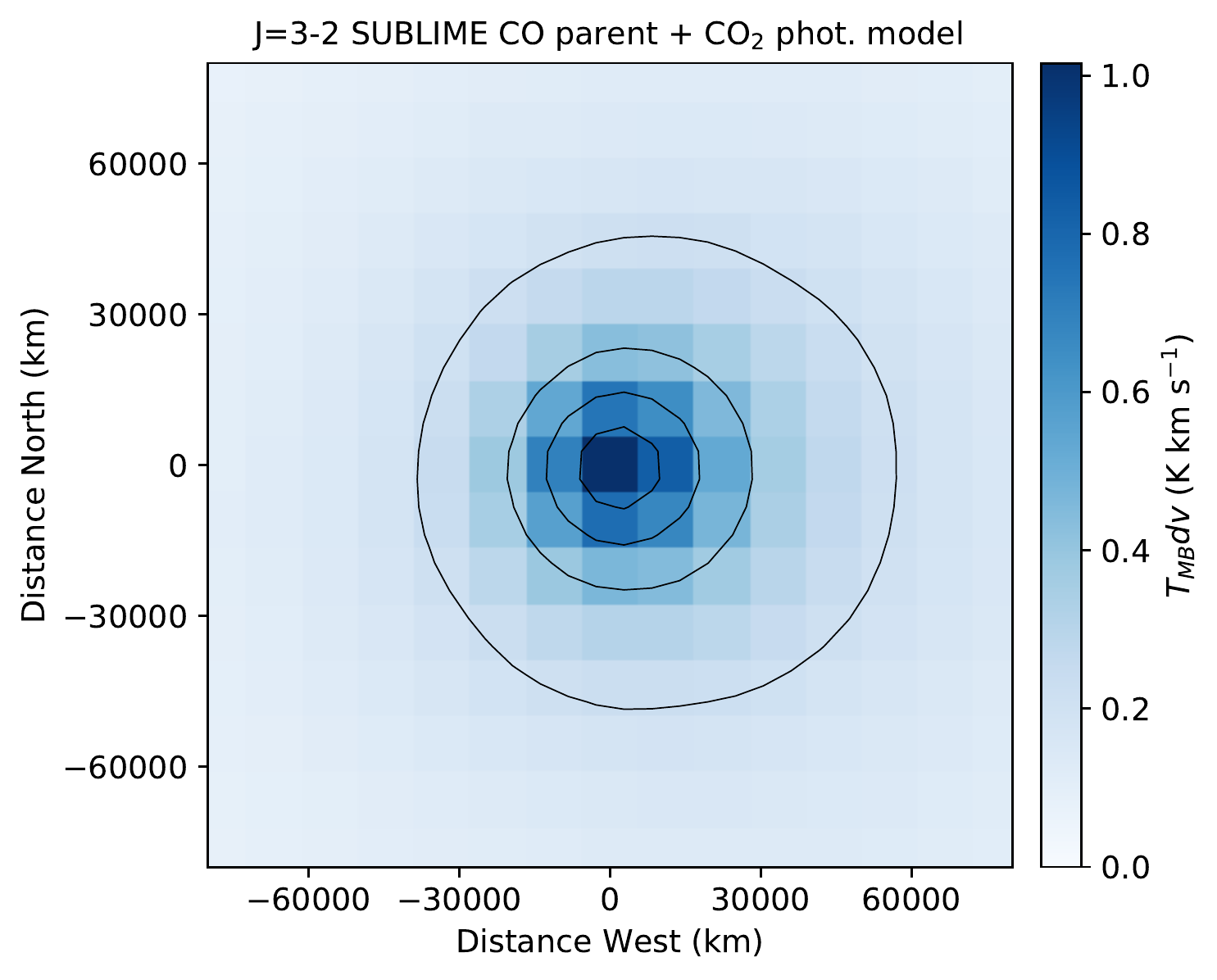} 
\includegraphics[width=0.32\textwidth]{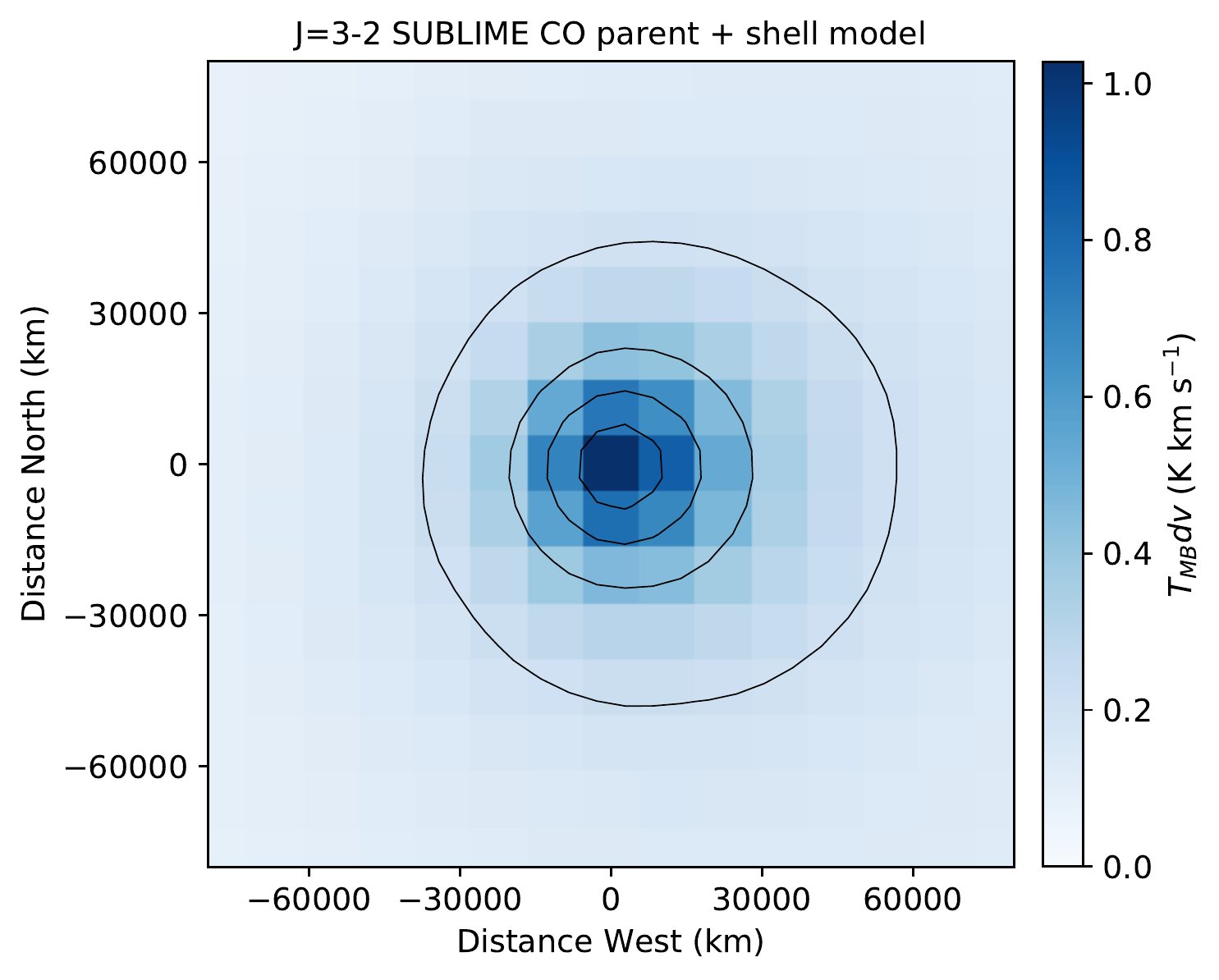}  
\caption{Spectrally integrated model CO $J=3-2$ maps (for comparison with Figure \ref{fig:jcmt_co_map}), assuming CO solely as a parent molecule (left panel), CO produced from the nucleus plus CO$_2$ photolysis, with $Q({\rm CO_2})/Q({\rm CO})=6.5$ (center panel), and CO from the nucleus, with additional CO shell at $r_c\sim1.4\times10^5$~km (right panel). Contour spacings are the same as in Figure \ref{fig:jcmt_co_map}. \label{fig:co_model_map}}
\end{figure*}

To investigate the spatial morphology of the CO coma, we generated 3D (spectral-spatial) models for comparison with the JCMT HARP image cube. First, a SUBLIME model fit was performed for the CO $3-2$ spectrum extracted from the central map pixel (coinciding with the inetnsity peak in Figure \ref{fig:jcmt_co_map}), by optimizing the model parameters $Q_1$, $Q_2$, $v_1$, $v_2$, $\theta$ and $\phi_{jet}$. The model orientation (in the plane of the sky) was fixed so that the jet azimuth angle matched the direction of the sky-projected comet-sun vector.  This best-fitting CO parent model was then convolved by the JCMT beam shape, multiplied by the main beam efficiency and spectrally integrated to obtain the 2D map in the left panel of Figure \ref{fig:co_model_map}.

\begin{figure}
\centering
\includegraphics[width=\columnwidth]{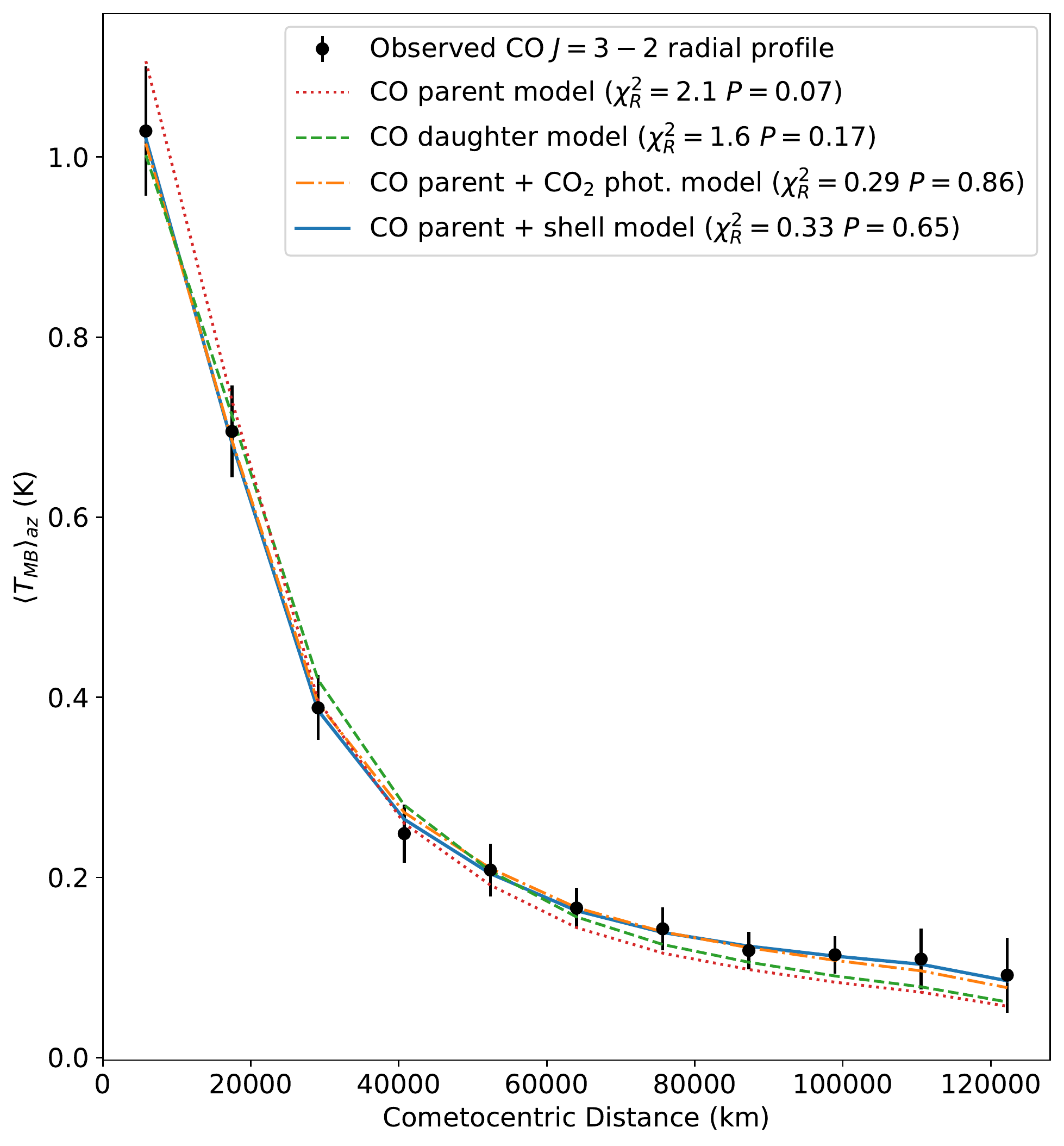} 
\caption{Azimuthal average (in 1-pixel bins) of the JCMT CO $J=3-2$ emission map (Figure \ref{fig:jcmt_co_map}) as a function of distance from the central pixel (black circles). Best-fitting SUBLIME model results are overlaid, including Haser-type parent and daughter models (see text). An improved fit is obtained at large radii when including a CO source from CO$_2$ photolysis (with a large $Q({\rm CO_2})/Q({\rm CO})$ ratio of 6.5), or with the inclusion of an extended CO shell at $r_s=1.2\times10^5$, with a density enhancement factor of $f=2.0$. \label{fig:azav}}
\end{figure}

Comparison of the contours in this figure with the observed CO map (Figure \ref{fig:jcmt_co_map}) shows that, while the central intensity peak is accurately reproduced, the outermost contour of the observations lies, on average, well outside that of the model. This is more clearly demonstrated by comparing the azimuthally averaged radial profiles ($\langle{T_{MB}}\rangle_{az}$) of the modeled and observed CO emission, which are plotted as a function of distance from the nucleus in Figure \ref{fig:azav}. The best-fitting parent model (green dotted line) falls off more rapidly with distance than the observed data (filled black circles), which implies the presence of an excess of CO emission at large radii, that cannot be explained solely by CO released from the nucleus at a constant outflow velocity.  The JCMT spectral baselines were well behaved over the course of our observations, and care was taken to ensure robust baseline removal through low-order polynomial subtraction, giving confidence regarding the reality of the observed extended CO emission (see also Figure \ref{fig:azavspec}). 

{ We quantify the statistical significance of the extended emission using the reduced chi-square statistic $\chi_R^2=\sum_i(y_i-y_m)/\sigma)^2/d$, where $y_i$ and $y_m$ are the observed and modeled intensities, respectively, and $d$ is the number of degrees of freedom (equal to the number of independent data points minus the number of free model parameters). The model $\chi_R^2$ value is given in the legend of Figure \ref{fig:azav}, along with the associated probability ($P$) that the difference between model and observations is due to statistical chance. With $\chi_R^2=2.1$ and $P=0.07$, the CO parent model does not represent a good fit to the observations}.

We consider three possibilities to account for the CO excess: (1) { a CO $J=3$ level population that increases more rapidly with distance than predicted}, (2) an additional Haser-type extended CO source, and (3) a step-like enhancement in the coma density at a large distance from the nucleus. A reduction in the CO photodissociation rate ($\Gamma_{\rm CO}$) would not reproduce the observed emission excess --- at $r_H=2.85$~au, $\Gamma_{\rm CO}=9.2\times10^{-8}$~s$^{-1}$ \citep{hue15}, so at an outflow velocity of $0.2$~km\,s$^{-1}$, the CO scale-length is $2.2\times10^6$~km. CO photodissociation is therefore negligible within the 80,000~km JCMT field of view, so reducing $\Gamma_{\rm CO}$ by a factor of a few (its range of uncertainty) has no noticeable impact on the modeled CO density profile.

\begin{figure}
\centering
\includegraphics[width=\columnwidth]{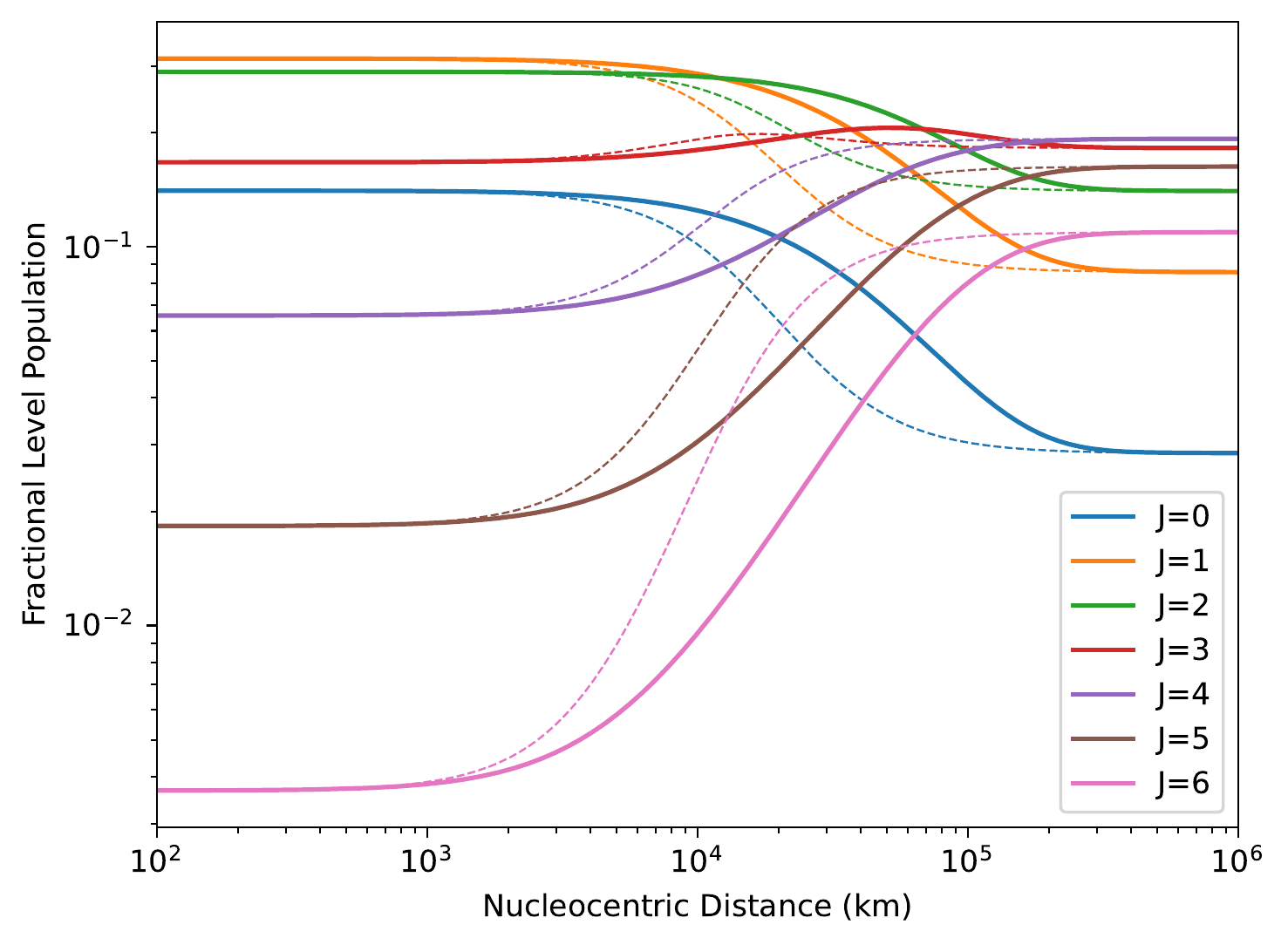} 
\caption{Fractional energy level populations as a function of radius for the lowest 7 CO rotational levels, using our SUBLIME time-dependent excitation model (solid curves) and the LIME steady-state solver (dotted curves). For clarity, only the populations from the higher-density jet component of the model (C$_1$) is shown. The steady-state model's failure to account for coma dynamics leads to the onset of non-LTE effects too close to the nucleus.  \label{fig:pops}}
\end{figure}

For option 1, we need to consider the CO rotational level populations. As shown by Figure \ref{fig:pops}, the population of the $J=3$ level (red curve) remains relatively constant as a function of distance from the nucleus. This is because the effective pumping rates ($G_{ij}$) into this level are closely balanced by radiative transitions out of the level, so on the larger distance scales $\gtrsim30,000$~km at which the excess CO emission becomes most visible, the $J=3$ level population is already close to the value attained at fluorescence equilibrium, making excitation effects an unlikely explanation for the observed extended emission. 

{ On smaller distance scales comparable with the size of the (central) JCMT beam, a lower rotational temperature could reduce the CO $3-2$ intensity, leading to a shallower radial profile in better agreement with the observations. However, assuming $T_{kin}=16.4$~K (the $1\sigma$ lower limit derived in Section \ref{jcmt_co_both}), the best-fitting $\chi_R^2$ value is 1.6, with $P=0.17$, which still does not represent a very good fit. Lower values of $T_{kin}$ also appear less likely given the range of possible $T_{kin}$ values (16--32~K) observed by \citet{biv18}, so we seek alternative explanations for the shape of the $\langle{T_{MB}}\rangle_{az}$ profile.}

The presence of an additional CO source (option 2) is worth considering, in light of previous evidence for extended CO sources in cometary comae \citep{cot08}. However, the JCMT map is { inconsistent} with CO produced solely as a (Haser-type) photochemical daughter species. {Our best fitting SUBLIME model was modified to produce CO from photodissociation of an (unknown) parent molecule, with the production rate ($Q_p$) and parent photodissociation rate $\Gamma_p$ free to vary (where $\Gamma_p=v_i/L_p$; $v_i$ is the outflow velocity in conical region $i$ and $L_p$ is the parent scale length)}. The best-fitting, azimuthally-averaged CO daughter model (with $\Gamma_p=7.3\times10^{-5}$~s$^{-1}$, $Q_p=7.0\times10^{28}$~s$^{-1}$) is shown in Figure \ref{fig:azav}, and systematically under-fits the observed CO emission at large radii (with $\chi_R^2=1.6$).

A better fit to the observations can be obtained using a composite CO density profile, where CO is included as both a parent and a daughter species. Figures \ref{fig:co_model_map} and \ref{fig:azav} show the best-fitting results from such a model, with CO originating from the nucleus at a total production rate $Q_t=5.5\times10^{28}$~s$^{-1}$, and an additional CO source in the coma from CO$_2$ photolysis (CO$_2$ + $h\nu \longrightarrow$ CO + O at $\Gamma_{\rm CO_2}=1.5\times10^{-7}$~s$^{-1}$; \citealt{hue15}). Although this composite model may appear plausible at first glance, and accurately reproduces the CO emission profile, it requires a CO$_2$ production rate 6.5 times larger than that of CO. Considering \citet{mck19} derived an upper limit of $Q({\rm CO_2})\lesssim1.5\times10^{28}$~s$^{-1}$ based on Spitzer observations from 2018-02-12 and 2018-02-21 (\emph{i.e.} 3.7 times less than the parent CO source), it seems unlikely that the comet could have been producing so much CO$_2$ at the time of our JCMT observations only a few weeks prior. Indeed, the large production rate required for any distributed molecular source to adequately fit the excess CO emission at large radii renders this an unlikely scenario in general, regardless of the assumed CO parent.

The possibility of non-Haser-type extended sources should also be considered, which could produce different radial density profiles, potentially resulting in a good fit to the observations. For example, \citet{gun02} developed a model for the sublimation and fragmentation of CO-rich icy grains to explain the extended CO distribution observed in comet 29P. Such detailed physical modeling is beyond the scope of our present study, however, but could be usefully investigated in a future article.

The third explanation we consider for the CO excess is an increase in coma density at large radii, which could result from temporal modulation of the gas production and/or outflow velocity, or a rapid onset of icy grain sublimation far from the nucleus. { Slowing of the gas expansion rate in the outer coma (for example, due to sublimation of slow-moving icy grains --- see \emph{e.g.} \citealt{ip86,fou12}), would also lead to an increase in gas density.} From Appendix \ref{sec:azavspec}, Figure \ref{fig:azavspec}, the CO line profile does not show any obvious evidence for coma deceleration with increasing distance, although we note that the weakness of the extended CO emission shell (representing $\sim1/6$ of the parent source emission at $r_c=60,000$~km) would make its velocity signature difficult to detect given the noise.

To model such an outer CO shell, we implemented a multiplicative increase in the coma density $n$ at radius $r_s$, smoothed by an exponential function such that $n(r) = 1+(f-1)/(1+e^{r_s-r}/w_s))$. The CO initial abundance, density enhancement factor ($f$), step radius ($r_s$) and smoothing width ($w_s$) were optimized to obtain the best fit to $\langle{T_{MB}}\rangle_{az}(r)$, and the resulting azimuthally-averaged model emission profile and moment 0 map are shown in Figures \ref{fig:azav} and \ref{fig:co_model_map}, respectively. The best-fitting model parameters are $f=1.8$, $r_s=1.2\times10^5$~km, $w_s=10^3$~km, { corresponding to $\chi_R^2=0.33$ and $P=0.65$}. The small value for $w_s$ implies that the density enhancement occurs abruptly, although larger values of $w_s$ (up to $\sim10^4$~km) also produce radial CO profiles consistent with the observations, within errors. The distance over which the implied density enhancement occurs is therefore not well constrained by our data. This is primarily because the radius of the shell (projected in the plane of the sky) lies just beyond the spatial extent of our JCMT map.

For an outflow velocity of 0.5~km\,s$^{-1}$, a factor of $\sim$ two density enhancement at $r_s=1.2\times10^5$~km would be consistent with a corresponding drop in $Q({\rm CO})$ around 67 hours earlier (on January 12th), but this seems perhaps unlikely given how uniform the comet's activity was over the weeks following that date (Figure \ref{fig:co_t}). The individual HARP maps from January 14th and 15th are also consistent with each other (within the $2\sigma$ noise level), and do not show any evidence for outward-moving CO density structure(s) over this period, so an onset of icy grain sublimation at large radii is our favored explanation.

Accounting for the extended CO shell leads to a marginally significant reduction in the nucleus production rate retrieved from the JCMT map, from $Q_t(\rm CO)=(6.1\pm0.8)\times10^{28}$~s$^{-1}$ to $(5.4\pm1.1)\times10^{28}$~s$^{-1}$. Adding the same extended CO component to our model for the time-averaged JCMT CO spectrum (Figure \ref{fig:co_avg}) does not significantly alter the quality of the spectral fit, but the resulting CO production rate from the nucleus ($(4.3\pm0.1)\times10^{28}$~s$^{-1}$) is 17\% lower.

\subsection{SMA CO Visibility Analysis}

Radio interferometry is a powerful technique for analyzing the radial distributions of cometary gases, to provide insight into their physical and chemical origins in the coma \citep{boi07,cor14,rot21}. This is due to the interferometer's ability to simultaneously sample emission from a range of spatial scales at high accuracy, from the near-nucleus environment to the outer coma. Interferometric maps (such as the SMA CO map in Figure \ref{fig:sma_co_map}), however, suffer from significant artifacts due primarily to the sparsely-filled telescope aperture, as well as Fourier image processing, re-gridding and deconvolution artifacts. Cometary coma images suffer in particular from a lack of information on the largest spatial scales ``missed'' by the interferometer. Consequently, the preferred method for robustly analyzing interferometric data of such extended sources is by directly modeling the calibrated `visibilities' recorded by the telescope (\emph{i.e.} the cross-correlation amplitudes between all antennas, as a function of baseline length).

\begin{figure}
\centering
\includegraphics[width=\columnwidth]{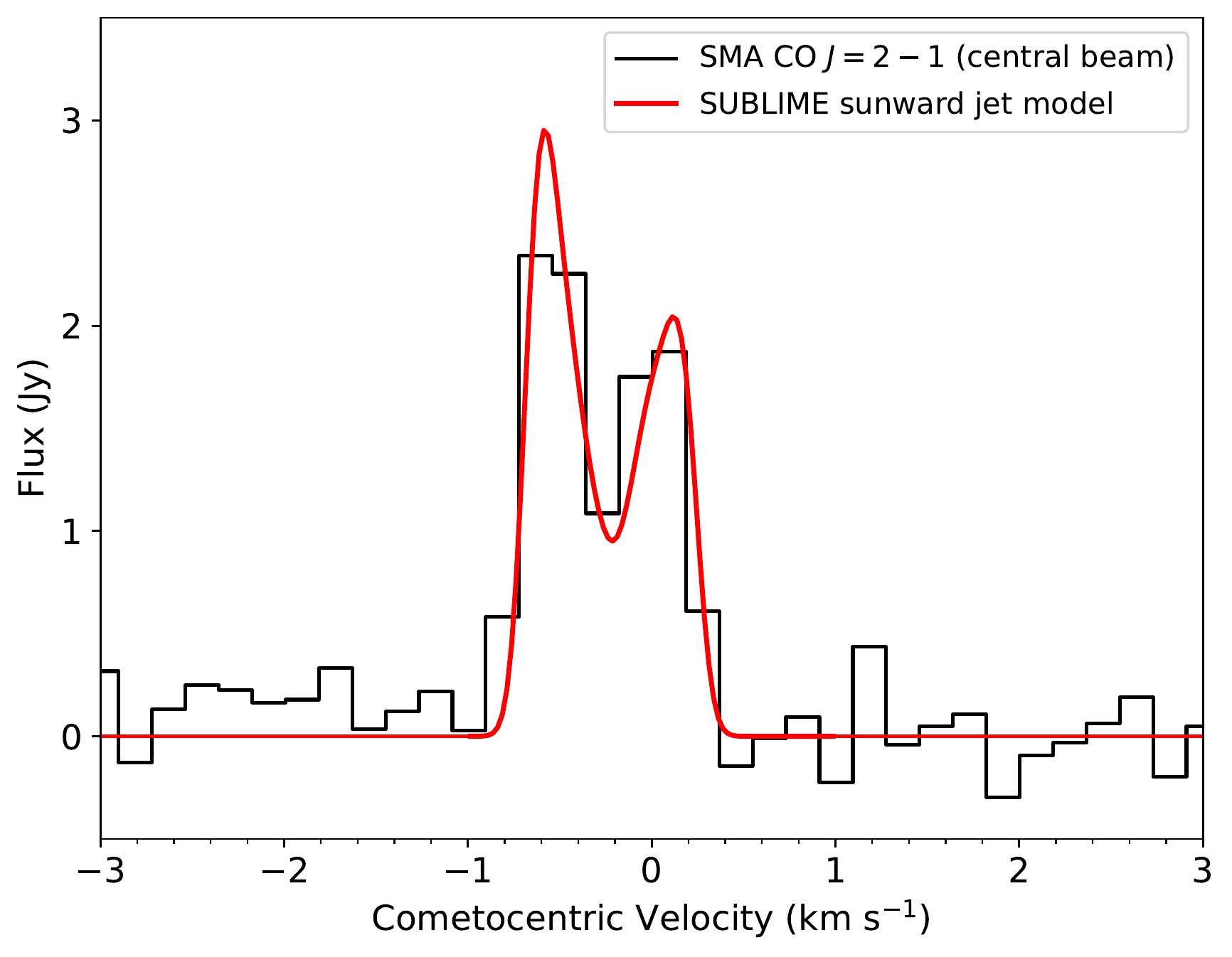} 
\caption{CO $J=2-1$ spectrum observed with the SMA (extracted at the CO emission peak), with best-fitting 2-component SUBLIME model overlaid using a red curve. \label{fig:sma_co_fit}}
\end{figure}

Visibility models were generated for the SMA CO $J=2-1$ observations of C/2016 R2, using our two-component SUBLIME model to test the same four scenarios as in Section \ref{sec:harp}. First, a fit was performed to the spectral line profile extracted at the emission peak of the SMA CO map (Figure \ref{fig:sma_co_map}). The best-fitting spectral line model is shown in Figure \ref{fig:sma_co_fit}, corresponding to $R_Q=9.2\pm1.5$, $v_1=0.64\pm0.02$~km\,s$^{-1}$, $v_2=0.20\pm0.02$~km\,s$^{-1}$ and $\theta=74^{\circ}\pm7^{\circ}$; the jet outflow axis was held fixed towards the sun due to a lack of constraints (resulting from the lower spectral resolution of this data). The different component outflow velocities compared to those derived from the JCMT observations 3-5 weeks earlier (Section \ref{sec:avg}) imply an increase in the jet outflow velocity accompanied by a slowing of the ambient coma. { This could be due to a stronger manifestation of intrinsic coma asymmetry at the higher spatial resolution of the SMA data compared with the JCMT}. Changes in CO outflow velocity could { also occur as a result of temporal activity variations} as the comet moved closer to the Sun.

\begin{figure}
\centering
\includegraphics[width=\columnwidth]{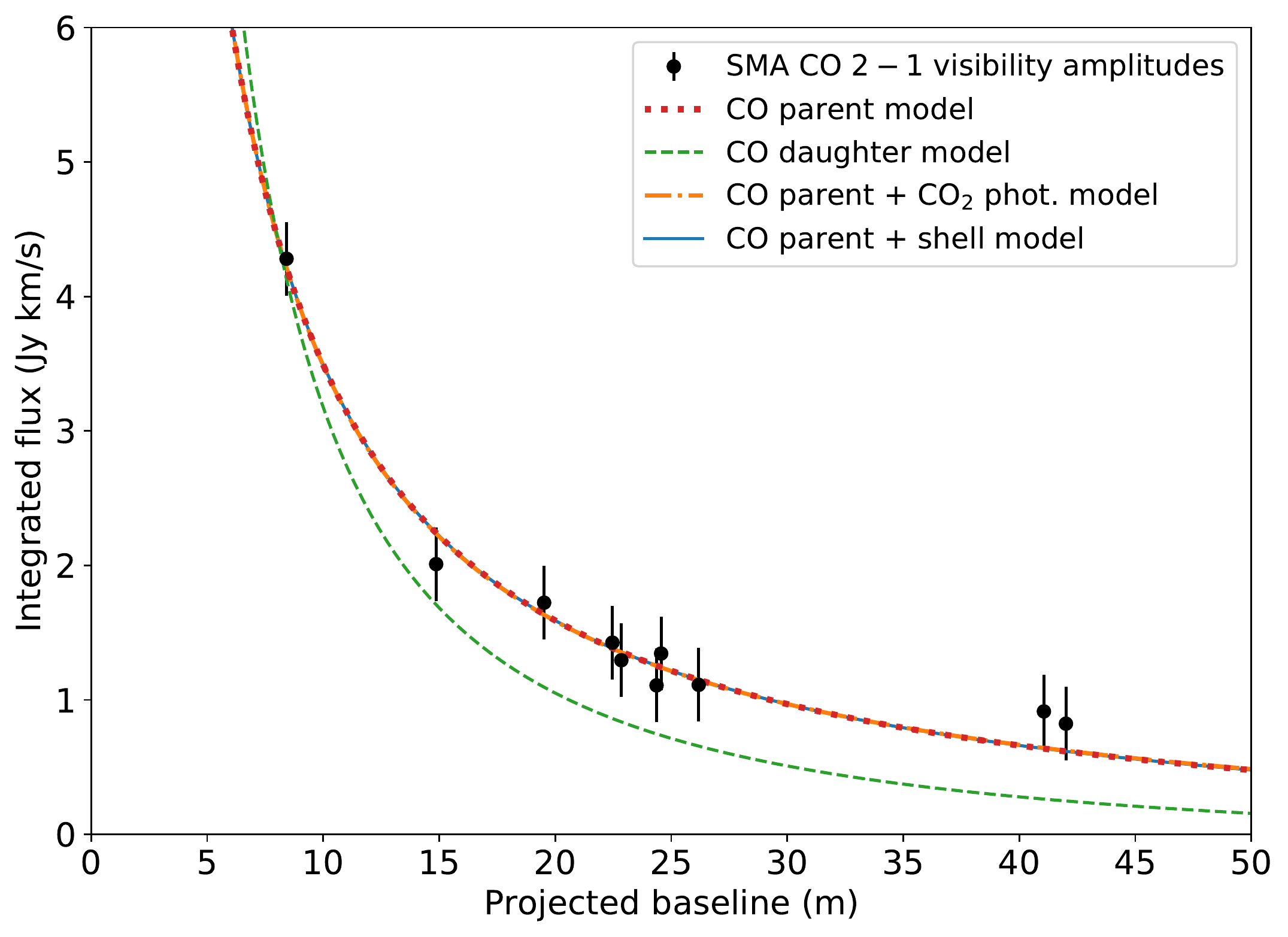} 
\caption{CO $J=2-1$ visibility amplitudes \emph{vs.} baseline length for C/2016 R2 observed using the SMA, including $1\sigma$ statistical error bars. Model visibility curves are overlaid for four different CO distributions. The CO parent, parent + CO$_2$ photolysis, and parent + shell model curves all lie on top of each other, whereas the CO daughter curve differs significantly, under-fitting the observations at large baseline (small angular scales). \label{fig:vis}}
\end{figure}

From this best-fitting `base' model, additional models were constructed assuming (1) CO solely as a daughter species, with $\Gamma_p=7.3\times10^{-5}$~s$^{-1}$ derived from the fit to the JCMT HARP data (Section \ref{sec:harp}), (2) CO as a parent with additional outer-coma source from CO$_2$ photolysis, and (3) CO as a parent with additional extended shell at $r_s=1.2\times10^5$~km, with $f=2.0$. The resulting 3D model images were integrated in the spectral domain then multiplied by the (FWHM $=55''$) SMA primary beam pattern before sampling in the Fourier domain using the {\tt vis\_sample} code \citep{loo18}. Visibility amplitude sampling was performed using the same set of $uv$ distances (baselines) as for the SMA observations (based on the time-averaged antenna positions during the comet observations). A power-law curve ($ax^b$) was fit through each set of model results, (with $x$ the baseline length, and $a$ and $b$ free parameters), and plotted along with the observed, time-averaged visibility amplitudes in Figure \ref{fig:vis}. For clarity, the model visibility curves were scaled vertically in order to pass through the shortest-baseline point. The CO production rate of the best-fitting (scaled) parent model is $Q_t(\rm CO)=(6.7\pm0.6)\times10^{28}$~s$^{-1}$ (the uncertainty increases to $0.9\times10^{28}$~s$^{-1}$ after inclusion of a 10\% amplitude calibration error). This value is $1.6\sigma$ from the mean CO production rate obtained using JCMT, corresponding to a barely-significant increase in $Q_t$ with time.

Three of the model curves (parent, parent + CO$_2$ photolysis, and parent + extended shell) all represent an equally good fit to the observations, falling precisely on top of each other in Figure \ref{fig:vis}. This implies that the SMA would have been blind to the additional CO component observed at large radii in the JCMT maps --- \emph{i.e.} the extended CO component was smooth enough on large angular scales not to be detected on even the shortest SMA baselines. In contrast, the CO daughter model does not fit the observed visibilities well, with insufficient flux on small angular scales (long baselines). Consequently, the SMA observations rule out the possibility of CO being solely a daughter species in this comet, although significant production at large radii from CO$_2$ photolysis is still possible.

\subsection{Other Molecules from JCMT: CH$_3$OH, HCN, H$_2$CO, $^{13}$CO and HCO$^+$}  
\label{sec:ul}

\begin{figure}
\centering
\includegraphics[width=\columnwidth]{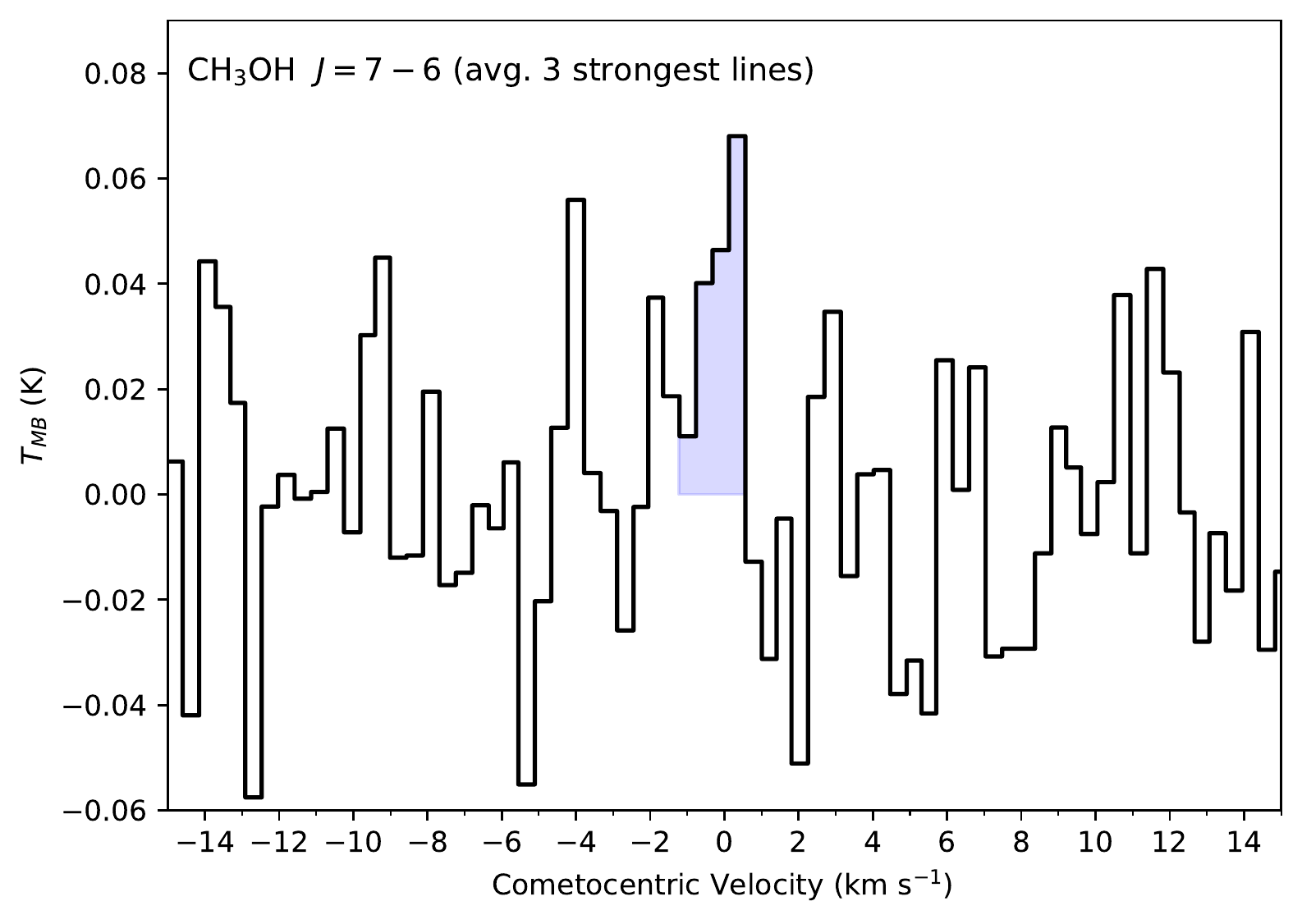} 
\caption{Average (in velocity space) of the three strongest CH$_3$OH transitions in our JCMT observations ($7_0-6_0\,E$, $7_{-1}-6_{-1}\, E$ and $7_0-6_0\, A^+$).  A tentative feature is present around 0~km\,s$^{-1}$ (the comet's rest velocity); blue shaded region shows the velocity range over which significant CO emission was detected. \label{fig:jcmt_ch3oh}}
\end{figure}

Our JCMT CH$_3$OH spectrum covered at least 12 lines of the $K=7-6$ band around 338 GHz (for details of the observed transitions, see \citealt{cor17b}, Table 1), but no individual CH$_3$OH lines were clearly detected in our data. We constructed a preliminary model for the CH$_3$OH spectrum based on the retrieved coma physical parameters from CO (Section \ref{sec:avg}), and identified three transitions that clearly stood out as stronger than the rest: $J_K=7_0-6_0\,E$, $7_{-1}-6_{-1}\, E$ and $7_0-6_0\, A^+$. The observed spectra for these three transitions were then averaged together in velocity space, producing the spectrum in Figure \ref{fig:jcmt_ch3oh}. A tentative feature is present around the comet's rest velocity (0~km\,s$^{-1}$), with a spectrally-integrated intensity (in the range $-1.0$ to $0.5$~km\,s$^{-1}$ shown by the blue shaded region) of $\int{T_{MB}dv}=73\pm20$~mK\,km\,s$^{-1}$, corresponding to a $3.6\sigma$ detection. Fitting the same three transitions simultaneously using SUBLIME (allowing only the CH$_3$OH abundance to vary) gave $Q({\rm CH_3OH})=(6.7\pm2.2)\times10^{26}$~s$^{-1}$, which corresponds to a CH$_3$OH/CO abundance ratio of $1.3\pm0.4$\% at the nucleus. For this model, we adopted CH$_3$OH--CO collisional transition rates based on the CH$_3$OH--H$_2$ rates from \citet{rab10}, with solar pumping rates from \citet{rot21b}. Uncertainties of a factor of five in the CH$_3$OH--CO collision rates lead to at most a 7\% error on the CH$_3$OH abundance.

\begin{figure}
\centering
\includegraphics[width=\columnwidth]{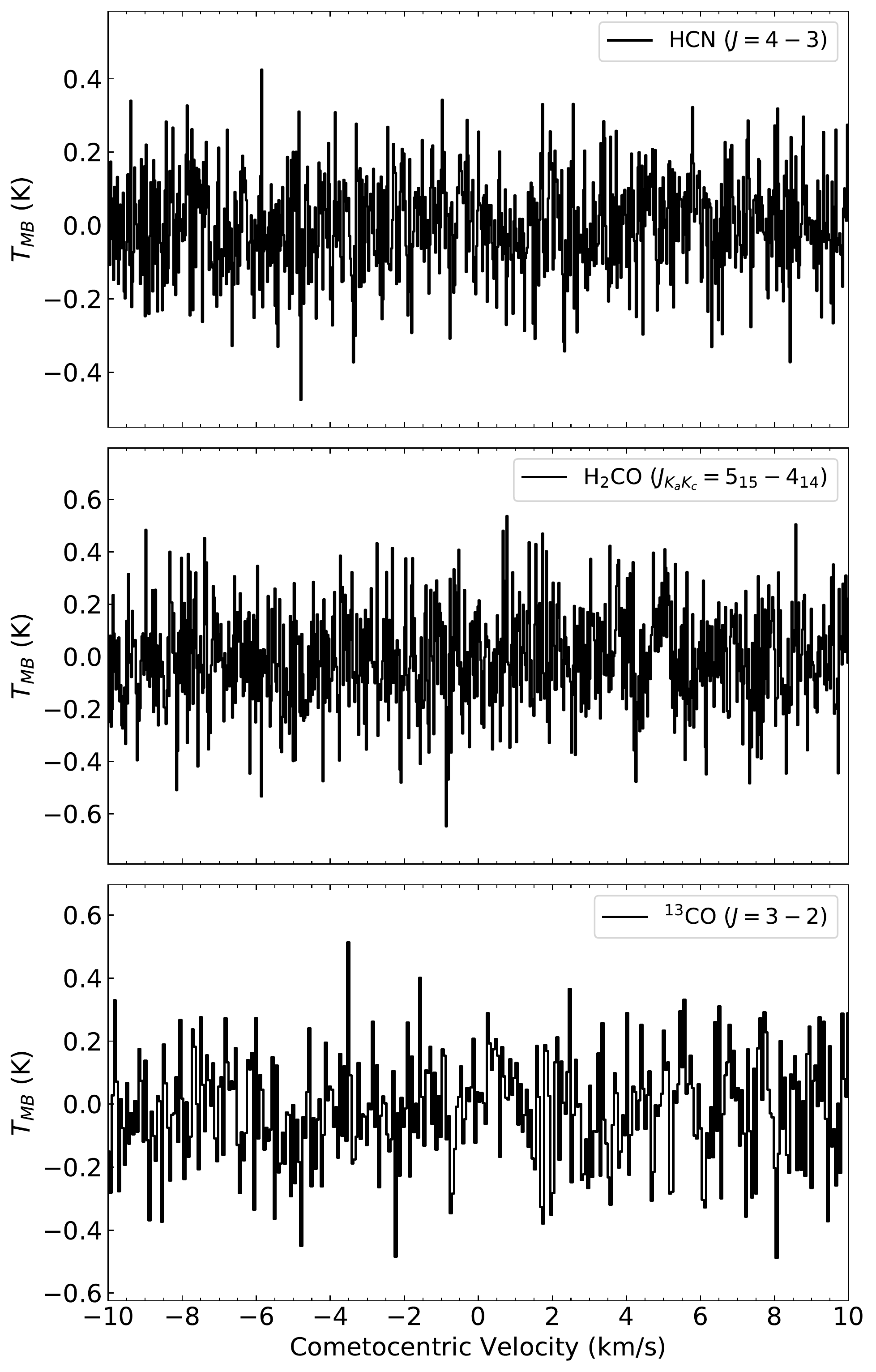} 
\caption{HCN, H$_2$CO and $^{13}$CO spectra observed using JCMT, showing no evidence for detections of these molecules. \label{fig:jcmt_ul}}
\end{figure}

The HCN, H$_2$CO and $^{13}$CO molecules were not detected in our data, as shown by the spectra in Figure \ref{fig:jcmt_ul}. Three-sigma upper limits were derived by comparing the spectrally-integrated noise level (from $-0.8$ to 0.5~km\,s$^{-1}$) to SUBLIME model line intensities for each molecule (again, using the same coma physical parameters derived in Section \ref{sec:avg}). For HCN we used the same collision and pumping rates as \citet{cor20}, for H$_2$CO we used the \citet{rot21b} rates, and for $^{13}$CO we used the same rates as for CO. Production rate $3\sigma$ upper limits are $Q({\rm HCN})<8.3\times10^{25}$~s$^{-1}$, $Q({\rm H_2CO})<2.3\times10^{26}$~s$^{-1}$ and $Q({\rm ^{13}CO})<2.7\times10^{27}$~s$^{-1}$, corresponding to abundance ratios HCN/CO $<0.16$\%, H$_2$CO/CO $<0.45$\% and CO/$^{13}$CO $>19$. The $^{12}$C/$^{13}$C ratio in CO is therefore consistent with the typical Solar System value of 89, as well as the value of $86\pm9$ measured in comet 67P \citep{alt19}.  

We also obtained a non-detection of HCO$^+$ $J=4-3$, with { $\int{T_{MB}dv}<164$~mK\,km\,s$^{-1}$ (integrated over the velocity range $\pm1.5$~km\,s$^{-1}$)}. This is { about an order of magnitude} less than the integrated HCO$^+$ $J=3-2$ line brightness observed in comet Hale-Bopp by \citet{mil04} using the SMT. However, the CO production rate in C/2016 R2 was also less than in Hale-Bopp by at least an order of magnitude at the time of observation \citep{biv97b}, so our result does not imply that C/2016 R2 had an unusually low HCO$^+$ production rate, { (considering the importance of CO in coma HCO$^+$ synthesis)}.

\section{Discussion}

\subsection{Comparison with Previous Observations of this Comet} 
\label{sec:compare}

Our primary result is the determination of a revised CO outgassing rate for C/2016 R2: $Q({\rm CO})=(5.3\pm0.6)\times10^{28}$~s$^{-1}$ (averaged over the JCMT observing period 2018-01-13 to 2018-02-01). This was subject to a marginally significant rise to $(6.7\pm0.9)\times10^{28}$~s$^{-1}$ on 2018-02-21, observed using the SMA. These results confirm C/2016 R2 as having among the highest CO production rates ever observed in a comet --- only a factor of 4 less than C/1995 O1 (Hale Bopp) at similar heliocentric distances.  Our CO production rates are slightly higher than the average value of $(4.6\pm0.4)\times10^{28}$~s$^{-1}$ between 2017-12-22 and 2018-01-16 observed by \citet{wie18} using the SMT, and similar to the $(5.5\pm0.9)\times10^{28}$~s$^{-1}$ observed on 2018-02-23 by \citet{mck19}, consistent with a slow, steady rise in CO activity as the comet moved closer to the Sun (between $r_H=2.98$--2.73~au). The observations by \citet{biv18} using the IRAM 30~m telescope on 2018-01-24 gave $Q({\rm CO})=(10.6\pm0.5)\times10^{28}$~s$^{-1}$, which is approximately double our JCMT value.  As shown in Appendix \ref{sec:bench}, the 1D versions of ours and Biver et al.'s models produce near-identical results. The discrepancy in CO production rates is therefore primarily attributable to the unusual complexity of the CO spectral line profile combined with differences between our 3D radiative transfer modeling strategies. 

{ To derive CO production rates, \citet{biv18} adopted a spherical coma model divided into three regions of `colatitude' ($\gamma$; angle measured from the Earth-comet vector), with outflow velocities $v=0.56$~km\,s$^{-1}$ between $\gamma=0$--60$^{\circ}$, $0.50$~km\,s$^{-1}$ between $\gamma=60$--120$^{\circ}$, and no outflow for $\gamma>120$.  Our model, on the other hand, gives $v_1=0.51$~km\,s$^{-1}$ in a sunward-facing conical jet and $v_2=0.25$~km\,s$^{-1}$ in the remaining (ambient) coma. Our chosen model geometry is physically justified based on fluid dynamic and Monte Carlo coma models \citep[\emph{e.g.}][]{cri99,fou16}, which show significantly different outflow velocities on the sunward and anti-sunward sides of the nucleus. The large reduction in the outflow velocity of our model across the entirety of the night side (most of which overlaps with the $\gamma>120^{\circ}$ region of no outflow in \citet{biv18}'s model) necessitates a lower total production rate in our model compared with that of \citet{biv18}.} 

{ Although our 2-component outflow model is sufficient to fit the data in the present study, we concede that it likely represents a simplification of the true physical situation.  Continuously variable $Q$ and $v$ as a function of (3D) coma position may be more correct, and could account for smoothly varying temperatures and mixing ratios within the nucleus, while avoiding discontinuities in the model parameters at the (conical) jet boundaries. The reality for C/2016 R2 may therefore involve a continuously variable $Q$ (as in the more complex model presented in Figure 3 of \citealt{biv18}), but with $v$ also allowed to vary more substantially/continuously, as may be expected based on physical models \citep[\emph{e.g.}][]{ten08}. Further contraints on the 3D outflow velocity distribution may be obtained from theoretical models, to help break the degeneracy that can occur between $Q$ and $v$ (particularly in the parts of the coma around $\gamma\sim90^{\circ}$), in order to obtain the most accurate production rates for an asymmetric coma.}

Compared to a completely spherically symmetric outflow model (as used by \citealt{wie18} and \citealt{mck19}), our best-fitting 2-component model for C/2016 R2 leads to { a $\sim30$\% lower CO production rate,} as a result of reduced outgassing on the anti-sunward side (see Figure \ref{fig:co_avg}). The exact correction factors are, however, dependent on the opening angle of the jet, and are larger if the jet is narrower. Based on the range of opening angles ($\theta=32$--$77^{\circ}$) obtained on our different JCMT observing dates (Table \ref{tab:fits}), correction factors in the range 29--48\% are possible, so the CO production rate reported by \citet{mck19} should be adjusted to $(2.9-3.9)\times10^{28}$~s$^{-1}$. Assuming { the comet's H$_2$O outgassing behaviour was similar to that of CO and CH$_3$OH (see \citealt{biv18} Figure 14), the \citet{mck19} H$_2$O production rate may also need to be adjusted. However, the correction factor is lower for OH (the H$_2$O daughter fragment observed by \citealt{mck19}), due to the OH kinetic energy gained in the photolysis reaction H$_2$O + $h\nu$ $\longrightarrow$ OH + H. Based on the outflow velocities from our best-fitting two-component CO parent model (Section \ref{sec:avg}), the vectorial model used by \citet{mck19} implies OH outflow velocities of 1.05~km\,s$^{-1}$ in the sunward jet and 0.78~km\,s$^{-1}$ in the ambient coma (M. Knight, private communication 2022). This geometry would correspond to a 22\% reduction in the H$_2$O production rate, assuming $\theta=62^{\circ}$.}

Abundance ratios from a given study tend to be less susceptible to model-dependent and instrumental uncertainties. Our CH$_3$OH/CO abundance ratio of $1.3\pm0.4$\% matches \citet{biv18}'s value of $1.04\pm0.08$\%, and our HCN/CO and H$_2$CO/CO upper limits are also consistent with their values of $(3.8\pm1.0)\times10^{-3}$\% and $0.043\pm0.006$\%, respectively. The CH$_3$OH/CO ratio is significantly higher than \citet{mck19}'s upper limit of 0.38\%, which could be indicative of significant temporal variability in the CH$_3$OH outgassing rate during 2018 January. Alternatively, the higher CH$_3$OH abundance observed using radio spectroscopy could have been due to additional CH$_3$OH production in the extended coma (from icy grain sublimation; \emph{e.g.} \citealt{cou17}), which was not detected on the smaller angular scales probed by the infrared observations of \citet{mck19}.

\subsection{Molecular Abundances and the True Nature of C/2016 R2 (PanSTARRS)}

After correcting the \citet{mck19} H$_2$O production rate for coma asymmetry (Section \ref{sec:compare}), we find a CO/H$_2$O ratio $\sim194$, which is $\sim3200$ times greater than the average value observed Oort cloud comets \citep{del16}, and $\sim42$ times greater than the highest previously observed in a comet (29P; \citealt{oot12}), thus confirming the extremely CO-rich nature of C/2016 R2's coma. The CO/H$_2$O abundance in cometary comae is observed to vary strongly as a function of heliocentric distance \citep[see][]{wie18,mck19}, due to the very different sublimation temperatures of these gases ($T_{sub}({\rm CO})=24$~K \emph{vs.} $T_{sub}({\rm H_2O})=152$~K), so a reduction in H$_2$O outgassing is expected for comets at heliocentric distances $\gtrsim3$~au, where the ice temperature falls below $T_{sub}({\rm H_2O})$. The CO/H$_2$O ratio in the coma of C/2016 R2 is several hundred to several thousand times greater than that found in other Oort cloud comets at similar heliocentric distances \citep{cro97,oot12,kaw14}, so this comet appears anomalous compared with all those previously observed. It should be kept in mind, however, that the coma abundances are not necessarily representative of those in the nucleus ices.

Cometary nuclei are heterogeneous, containing mixtures of ices in different phases and compositions \citep{mum11,ahe11,alt19}. Chemical differentiation as a function distance below the surface, or the presence of a volatile-depleted outer crust \citep[\emph{e.g.}][]{cap15} could lead to H$_2$O being insulated from solar heating while CO continues to sublimate. For example, the 3D numerical nucleus model of \citet{mar14} demonstrates that the ratio of CO to H$_2$O production rates could be enhanced by several orders of magnitude by the presence of an insulating dust mantle $\sim5$--10~cm thick that hinders H$_2$O sublimation. Alternatively, a moderate overabundance of CO ice close to the surface could lead to increased cooling by CO sublimation, helping keep the nucleus at { a low enough temperature to inhibit H$_2$O sublimation} \citep{lis21}. In that case the comet could maintain a lower H$_2$O outgassing rate (relative to CO) for a longer duration as it approached the sun. Either case may not require an extremely anomalous CO/H$_2$O abundance in the bulk nucleus.

Peculiarities in the coma abundance ratios for several other molecules were reported by \citet{biv18} and \citet{mck19}. Despite similar sublimation temperatures for CH$_3$OH and HCN (99~K and 95~K, respectively), these two molecules were enriched in the coma (relative to H$_2$O) by very different amounts: the CH$_3$OH/H$_2$O ratio was 163 times the Oort cloud comet average, whereas HCN/H$_2$O was only 5.3 times the average. Evidently, such enrichment patterns cannot be produced by reduced H$_2$O outgassing alone, or by a simple temperature-dependence of the CH$_3$OH and HCN sublimation rates. Similarly, the coma CH$_4$/H$_2$O ratio is a factor of 206 higher than average, but this is several hundred times less than the CO/H$_2$O enrichment, despite similar sublimation temperatures for these two molecules (31~K and 24~K, respectively). It is tempting to take such unusual coma abundance patterns to be directly representative of a peculiar nucleus ice composition, but before doing so, it is worth considering the possible role played by ice heterogeneity and molecular trapping at temperatures below $T_{sub}({\rm H_2O})$ (\emph{i.e.}, with the comet not yet fully activated). 

As shown by laboratory ice sublimation experiments \citep{col04}, volatile gases can be trapped in mixed (H$_2$O-dominated) ices at temperatures well above their sublimation points, and may be only partially released until the H$_2$O sublimation temperature is reached ($\sim 150$~K). Release of trapped hypervolatiles can also occur as H$_2$O ice undergoes a phase change from amorphous to crystalline (at $T_{AC}\sim130$~K). Such trapping and release processes are inevitable in mixed cometary ices, and could explain some of the observed abundance patterns in C/2016 R2. Closely related outgassing behaviors were observed for CO, CO$_2$ and CH$_3$OH in comet 67P \citep{lau19,biv19}, consistent with our understanding regarding a common origin for these species in carbon- and oxygen-rich interstellar ices \citep{fuc09,iop11,gar11}. Observations of young stellar objects also indicate mixing of CO, CO$_2$ and CH$_3$OH ices in an apolar phase, distinct from the polar, H$_2$O-dominated ice \citep{boo15,pen15}. In-situ observations of comets 103P \citep{ahe11} and 67P \citep[\emph{e.g.}][]{mig16,gas17} discovered spatial {and temporal variations} in the coma H$_2$O/CO$_2$ ratios, implying non-uniform mixing ratios for these two volatiles within the nucleus. {As shown by \citet{dav21,dav22} for 67P, this could be caused by physical/chemical evolution of the comet's surface layers due to anisotropic illumination of the comet, but primordial variations in the H$_2$O/CO$_2$ ratio intrinsic to the nucleus are also possible}. It is therefore plausible that CH$_3$OH in C/2016 R2 exists as a component within an apolar ice matrix dominated by CO and CO$_2$, and some of this CH$_3$OH is released into the coma when CO sublimates (as observed). Meanwhile, a significant CH$_3$OH component also likely remains frozen as part of the comet's polar, H$_2$O-rich ices \citep{qas18}. HCN, on the other hand, could be primarily associated with the (still frozen) H$_2$O ice rather than the (sublimating) CO component, and would then be only partially outgassed at $T_{sub}({\rm HCN})$ or $T_{AC}$. Future studies of cometary CO, CH$_3$OH and HCN spatial distributions could help test this hypothesis.

To complete this picture, the C$_2$H$_6$ and NH$_3$ upper limits from \citet{mck19} are consistent with moderate-to-no enrichment (with respect to H$_2$O), and could therefore be associated primarily with the H$_2$O-dominated ice, while the strongly-enriched N$_2$ and CO$_2$ are associated more with CO. Observations and laboratory studies show that interstellar CH$_4$ ice tends to be more associated with H$_2$O than CO \citep{obe08,qas20}, while the location of H$_2$CO ice is less well constrained. The CH$_4$ and H$_2$CO enrichment factors in C/2016 R2 (206 and 43, respectively) could thus be explained by release of trapped volatiles (in H$_2$O ice) above their respective sublimation temperatures (31~K and 64~K) \citep{col04}. The {increasing similarity to typical} abundances (with respect to H$_2$O) in the sequence CH$_4$ $\to$ H$_2$CO $\to$ HCN is consistent with the decreasing volatility (increasing $T_{sub}$) of these three species, such that they each behave progressively more like H$_2$O. We therefore postulate the existence of a rapidly sublimating (apolar) component of ice in C/2016 R2 rich in CO, CO$_2$, N$_2$ and CH$_3$OH, and a second (polar) component containing more CH$_4$, H$_2$CO and HCN mixed with H$_2$O. If the N$_2$/CO ratio is higher in the CO-rich ice phase than the H$_2$O-rich ice, this could explain why the N$_2$/CO ratio is lower in fully-activated comets than in C/2016 R2.

\subsection{Coma Morphology}

The highly asymmetric, blueshifted CO line profile of C/2016 R2 is similar to that of the large Centaur 29P/SW 1 \citep{fes01,gun02}, as well as to the CO line profile observed in C/1995 O1 at $r_H\gtrsim8$~au \citep{gun03b}. Our interpretation of the line shape in terms of enhanced CO production and outflow velocity on the sunward side of the nucleus is consistent with \citet{gun08}'s analysis of the 29P coma, and we find a similar ratio of day-to-night hemisphere $Q({\rm CO})$ and $v_{out}$ values in C/2016 R2.

\citet{gun02} also discovered an extended shell of CO emission surrounding 29P, at a cometocentric distance $r_c\sim1.4\times10^5$~km, based on mapping observations of the $J=2-1$ line in 1998 (although the shell was no longer apparent in 2003 followup observations, demonstrating an intermittent nature; \citealt{gun08}). The spatial properties of this shell are remarkably similar to those of the extended CO emission structure we detected at a similar cometocentric distance in our C/2016 R2 JCMT maps (Section \ref{sec:harp}). \citet{gun03} interpreted the extended CO structure in 29P as arising from sublimation of a population of icy grains long-lived enough to reach 29P's outer coma. Such an explanation also appears plausible for C/2016 R2, although more detailed modeling would be required to confirm this possibility, and to test the other possible origins for the shell considered in Section \ref{sec:harp}. We do not consider CO$_2$ to be a likely source for the majority of the extended CO in C/2016 R2 as it would require at least an order of magnitude larger CO$_2$ production rate than found by \citet{mck19}.

A time-variable, diffuse, ring-shaped feature (consistent with excess gas emission) also appeared in Spitzer IRAC CO + CO$_2$ images of C/2016 R2 on 2018-02-21 --- a feature that was not apparent 9 days earlier (see Figure 5 of \citealt{mck19}). The approximate diameter of this ring was also $\sim1.4\times10^5$~km (M. Kelley, private communication), so it could plausibly be related to the shell-like feature observed using JCMT. Detailed modeling of the Spitzer images will be required to determine the physical properties of this ring, and to confirm whether icy grain sublimation or coma deceleration could be responsible.

On both dates, the Spitzer images also show enhanced gas emission within an angular wedge (of opening angle $\sim80^{\circ}$), oriented toward the (sky-projected) comet-sun vector, with a morphology consistent with a jet or fan emanating from the nucleus. Although this feature may be attributable to a combination of both CO and CO$_2$ emission, its apparent qualitative similarity to the jet properties derived from modeling our JCMT and SMA data provides further evidence for preferential sunward outgassing from a confined region of the nucleus. This provides additional validation of our 2-component coma model.


\subsection{Uncertainties in the CO Excitation}
\label{sec:unco}

\begin{figure}
\centering
\includegraphics[width=\columnwidth]{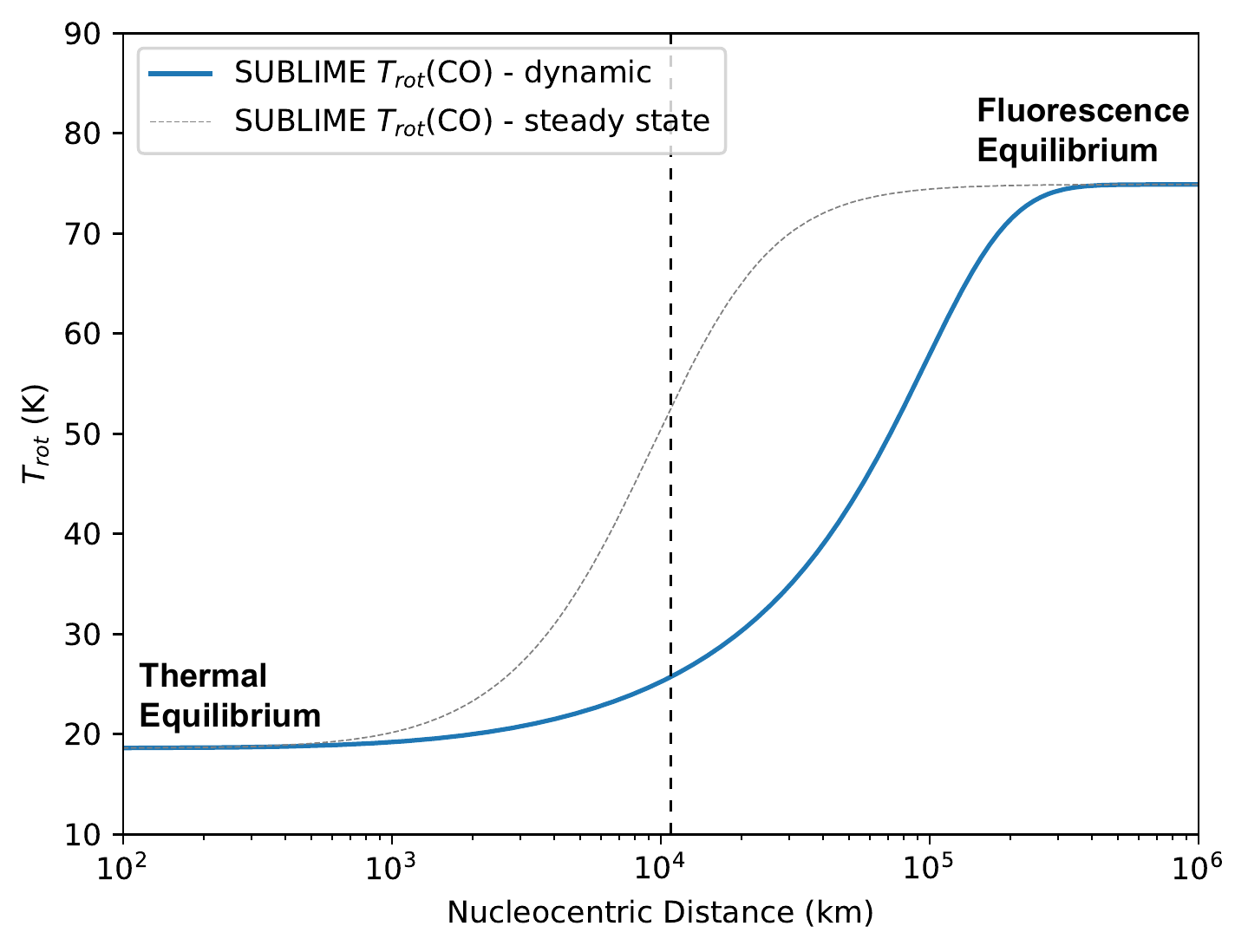} 
\caption{Modeled CO rotational temperature derived using the 7 lowest energy level levels, from our time-dependent SUBLIME model (solid blue curve) and the steady-state version of the model (dotted grey curve), based on the data shown in Figure \ref{fig:pops}. The rotational temperature evolves from thermal equilibrium at the gas kinetic temperature (18.7~K) in the collisionally-dominated inner coma, to fluorescence equilibrium (at 75 K) in the solar radiation-dominated outer coma. The vertical dashed line shows the radial extent of the $14''$ JCMT beam FWHM at the Geocentric distance of C/2016 R2 (2.14~au), at 346~GHz. \label{fig:trot}}
\end{figure}

Our new CO--CO collisional rate coefficients are considered to be accurate to within about a factor of 2 (Section \ref{sec:rates}). As shown in Figure \ref{fig:bench} (Appendix \ref{sec:bench}), differences in the rates on that order lead only to relatively small discrepancies in the CO rotational level populations, the evolution of which is controlled to a large extent by radiative processes, especially in the outer coma. As a result, varying the collision rates by $\pm$ a factor of two leads to changes of only $\pm1$\% in our retrieved CO production rates.

An additional source of error in the CO excitation calculation arises from the fact that the collision rates are tabulated as a function of $T_{rot}$, which provides only an average measure of the true distribution of energy level populations. In a non-LTE regime, individual level populations may deviate from the Boltzmann distribution, in which case the (thermally averaged) state-to-state collision rate coefficients are no longer accurate. Perhaps more importantly, the rate coefficients are dependent on $T_{rot}$ as well as $T_{kin}$, yet for the purpose of choosing rates in our model, we assume $T_{rot}=T_{kin}$, which is only strictly accurate in the collisionally-dominated inner coma. Both these issues have negligible impact on the results of our study, however, since the collision rates $k_{ij}$ are found not to vary strongly as a function of $T_{rot}$, changing by on average only 9\% between 10--30~K.

For deriving accurate CO abundances, rather than having the most precise set of collision rate coefficients, it is more important to use the most physically accurate, time-dependent solution to the equation of statistical equilibrium \citep{boc87,biv97}, as opposed to the steady-state approximation \citep{hog09,val18}. Figure \ref{fig:pops} shows a comparison between the CO level populations as a function of radius using these two different methods, while Figure \ref{fig:trot} shows the corresponding rotational temperature behaviour. As described by \citet{gar20}, species with small dipole moments such as CO have long timescales $\tau_R$ with respect to spontaneous (and induced) rovibrational radiative transitions. If $\tau_R$ is greater than the dynamical timescale in the outflow (\emph{i.e.} the time for the coma gas to move a given distance $r_d$), then an accurate calculation of the excitation over distances $r_d$ requires the gas motion to be considered, and the steady-state approximation is no longer valid.

An observational consequence for molecules with large $\tau_R$ is that the non-LTE effects occur further out in the coma, because the time-dependent solution (solid curves in Figures \ref{fig:pops} and \ref{fig:trot}) is `delayed' with respect to the steady-state solution (dashed curves). For the $J=3$ level populations, the discrepancy is minimal (amounting to $<1$\% difference in the $J=3-2$ line integrated intensity). However, for the $J=2-1$ line, the steady-state solution overestimates the line intensity by 19\%. For larger beam sizes more sensitive to the non-LTE region of the coma (between the thermal and fluorescence equilibrium extremes highlighted in Figure \ref{fig:trot}), the discrepancy can be even larger. Consequently, the use of a time-dependent excitation model is recommended for correct analysis of single-dish cometary CO data. It should be born in mind that this issue is less severe for molecules with larger dipole moments (and therefore, smaller $\tau_R$ values) such as H$_2$O, HCN, CH$_3$OH and H$_2$CO, although in low-activity comets, the premature departure from LTE of the steady-state solution can still lead to some large discrepancies. For example, in a comet at $r_H=\Delta=1$~au, with $Q({\rm H_2O})=10^{27}$~s$^{-1}$, $T_{kin}=50$~K and $v_{out}=0.8$~km\,s$^{-1}$, the HCN $J=4-3$ line intensity (for a $14''$ JCMT beam) is overestimated in the steady-state model by 46\%, whereas at $Q({\rm H_2O})=10^{29}$~s$^{-1}$, the discrepancy is reduced to only 4\%.

\section{Conclusions}

Based on high-resolution spectral-spatial observations using the JCMT and SMA telescopes during the period 2018-01-13 to 2018-02-21, we confirm the presence of extremely strong, asymmetric CO outgassing from the hypervolatile-rich comet C/2016 R2 (PanSTARRS), with $Q({\rm CO})$ in the range $(3.8-7.6)\times10^{28}$~s$^{-1}$. The observational data were analyzed using a new, time-dependent, 3D radiative transfer code, adopting a two-component model for the expanding coma, for the first time using state-to-state CO--CO collision rates based on quantum scattering calculations using the coupled-states method. We determined the presence of a (near-)sunward CO jet with a (time-variable) half-opening angle of $\theta=25$--$90^{\circ}$ (and an average of $\theta=62\pm2^{\circ}$, offset by $7\pm2^{\circ}$ from the sun-comet vector). The average jet outflow velocity determined from our JCMT data was $0.51\pm0.01$~km\,s$^{-1}$, while the ambient coma outflow velocity was found to be $0.25\pm0.01$~km\,s$^{-1}$. The total amount of CO produced by the jet was, on average, a factor of 2.0 times more than that of the ambient coma.

On 2018 January 14-15, { we found evidence for extended CO emission that cannot be easily explained by standard nucleus outgassing or excitation effects. The extended emission is therefore interpreted as a possible} result of modulation in the CO outgassing rate, deceleration in the outer coma, or sublimation of long-lived icy grains. Subtraction of such an extended CO component from the best-fitting (time averaged) JCMT $3-2$ model results in a 17\% reduction in the CO production rate derived for the nucleus. Using a value for $Q({\rm H_2O})$ from \citet{mck19}, corrected for asymmetric outgassing, our CO/H$_2$O ratio is $\sim42$ times larger than seen in any comet to-date (including the distant Centaur 29P), which, combined with previously-noted chemical peculiarities, suggests C/2016 R2 is among the most unusual comets ever observed.

However, the heterogeneous nature of cometary ices, combined with knowledge that molecular outgassing rates from mixed ices do not necessarily correlate with their sublimation temperatures means that we cannot yet rule out a bulk composition for C/2016 R2 more similar to the general population of Oort cloud comets than previously inferred. We hypothesize that the ice temperature of C/2016 R2 may have been suppressed by sublimative cooling, or the presence of an unusually thick insulating crust, that prevented the initiation of a more conventional, H$_2$O-dominated outgassing regime, leading to sublimation rates more heavily influenced by trapping and binding of individual molecules within the bulk ices. We propose that the observed abundance patterns can be explained by the existence of two ice phases in C/2016 R2, similar to those observed in young stellar objects: (1) an apolar phase, rich in CO, CO$_2$, N$_2$ as well as CH$_3$OH ices, and (2) a polar phase containing larger abundances of CH$_4$, H$_2$CO and HCN mixed in with H$_2$O ice. More observations of coma chemistry in distant comets (at $r_H\gtrsim2.5$ au, for which H$_2$O sublimation is not yet fully activated), will be crucial to better understand this comet's peculiar nature, and to constrain the physical and chemical processes that govern the formation, storage and release of cometary volatiles below the H$_2$O sublimation point.

\acknowledgments
This work was supported by the National Science Foundation under Grant No. AST-2009253. The work of SNM, NXR, MAC, EGB and SBC was also supported by NASA’s Planetary Science Division Internal Scientist Funding Program through the Fundamental Laboratory Research work package (FLaRe). SNM and SBC were supported by the NASA Astrobiology Institute through the Goddard Center for Astrobiology. The James Clerk Maxwell Telescope is operated by the East Asian Observatory on behalf of The National Astronomical Observatory of Japan, Academia Sinica Institute of Astronomy and Astrophysics, the Korea Astronomy and Space Science Institute, the National Astronomical Observatories of China, and the Chinese Academy of Sciences (grant No. XDB09000000), with additional funding support from the Science and Technology Facilities Council of the United Kingdom and participating universities in the United Kingdom and Canada. The Submillimeter Array is a joint project between the Smithsonian Astrophysical Observatory and the Academia Sinica Institute of Astronomy and Astrophysics and is funded by the Smithsonian Institution and the Academia Sinica. We gratefully acknowledge the assistance of R. Loomis and S. Andrews with the use of the {\tt vis\_sample} code. We thank N. Biver for providing comparison radiative transfer model calculations. Thanks also to A. McKay for providing comments on the manuscript regarding the nucleus and coma composition.

\software{CASA (v5.1; \citealt{jae08}), SMA-MIR (https://github.com/qi-molecules/sma-mir), LIME code \citep{bri10}, CVODE solver \citep{hin19}, Moslcat \citep{mol94}, vis\_sample \citep{loo18}, MPFIT \citep{mar12}.}

\appendix

\section{Equations of Radiative Transfer and Molecular Excitation}
\label{sec:radtran}

The intensity of radiation ($I_{\nu}$) propagating through the coma at a frequency $\nu$ is calculated by integrating the equation of radiative transfer as a function of distance along the line of sight ($s$):

\begin{equation}
\label{eq:radtran}
\frac{dI_\nu}{ds} = j_\nu - \alpha_\nu I_\nu,
\end{equation}
where $j_\nu$ and $\alpha_\nu$ are the gas emission and absorption coefficients, respectively. These are derived from the Einstein coefficients of the gas ($A_{ij}$, $B_{ij}$ and $B_{ji}$, for a transition between the upper energy level $i$ and lower level $j$), as follows:

\begin{equation}
j_\nu = \frac{h\nu}{4\pi}N_jA_{ij}\psi_\nu,
\end{equation}

\begin{equation}
\alpha_\nu = \frac{h\nu}{4\pi}(N_iB_{ij}-N_jB_{ji})\psi_\nu
\end{equation}

where $N_i$, $N_j$ are the number of gas particles per unit volume in levels $i$ and $j$, respectively, and $\psi_\nu$ is a (normalized) line-broadening function for the spectral line of interest (typically a Gaussian for individual, thermally-broadened lines).

The number of molecules (per unit volume) in energy level $i$ is obtained as a function of time ($t$) in the outflowing coma gas by solving the following differential equation (\emph{e.g.} \citealt{cro87}):

\begin{equation}
\begin{split}
	\frac{dN_i}{dt} = -N_i \left[\sum_{j<i} A_{ij} + \sum_{j\neq i} (B_{ij}J_\nu + k_{ij}n + G_{ij})\right]\\
	+ \sum_{j>i}N_j A_{ji} + \sum_{j \neq i}N_j(B_{ji}J_\nu + k_{ji}n + G_{ji}).
\end{split}
\label{eq:excite}
\end{equation}

In this equation, $k_{ij}$ are the rates at which transitions occur between { rotational} levels $i$ and $j$ { (in the ground vibrational state)} due to collisions between the gas particles, $n$ is the number density of colliders (in this case, the CO gas density), and $G_{ij}$ are the effective transition rates due to fluorescence / vibrational pumping by solar radiation, summed over the relevant rovibrational bands (see \emph{e.g.} \citealt{cro83,ben04}).  The energy level populations at a given location in the coma depend on the local radiation field, $J_{\nu}$, which is calculated by summing the incident radiant energy received at that point from all solid angles. In general, this means that equation \ref{eq:excite} needs to be solved iteratively until convergence of $J_{\nu}$ is achieved. In practice however, { for species other than H$_2$O} the optical depth for photons leaving the less dense parts of the coma where non-LTE effects are important tends to be low (\emph{i.e.} $\tau_\nu\ll1$). In that case, the stimulated emission and absorption terms ($B_{ji}J_{\nu}$) are small, and can be neglected. For more optically-thick gases, the escape probability method can be used as a quick (and easy to implement) approximation for the effects of photon trapping \citep[\emph{e.g.}][]{boc87}.

Solar radiation-induced fluorescence (pumping) is responsible for modifying the rotational level populations. Effective pumping rates ($G_{ij}$) for CO were calculated using the method of \citet{cro83}, incorporating the latest infrared transition data from the HITRAN catalog \citep{gor21}. The effective pumping rates were summed over all rovibrational transitions involving the ground vibrational state of CO. Excitation of the gases of interest due to collisions with coma electrons has also been implemented in SUBLIME using the method of \citet{biv97} and \citet{zak07}, but for the present study focusing on CO emission from a CO-dominated coma, the electron-collision rates are found to be small enough that they can be neglected.

\section{State-to-state collision rates for the CO--CO system}
\label{sec:ratestab}

Table \ref{tab:rates} shows collisional (de-excitation) rate coefficients $k_{J_1 \to J_1'}=k_{ij}$ (in cm$^{-3}$\,s$^{-1}$) for gas-phase CO molecules undergoing transitions $J_1 \to J_2$, as a function of kinetic { temperature} ($T_{kin}$). The reverse (excitation) rates are calculated from the principle of detailed balance according to Equation 6 of \citet{van07}. For further details of the rate coefficient calculations, see Section \ref{sec:rates}.

\begin{table*}
\centering
\caption{CO--CO state-to-state collision rate coefficients $k_{J_1 \to J_1'}$ as a function of $T_{kin}$ \label{tab:rates}}
\begin{tabular}{llcccc}
\hline
\hline
&&\multicolumn{4}{c}{$T_{kin}$}\\
\cline{3-6}
$J_1$&$J_1'$&5 K&10 K& 20 K& 30 K\\
\hline
 1 &  0  &   2.69e-11& 3.61e-11& 4.01e-11& 4.13e-11 \\
 2 &  0  &   2.73e-11& 3.03e-11& 2.97e-11& 2.81e-11 \\
 2 &  1  &   5.56e-11& 6.23e-11& 6.69e-11& 6.70e-11 \\
 3 &  0  &   1.81e-11& 2.02e-11& 1.96e-11& 1.81e-11 \\
 3 &  1  &   4.73e-11& 5.14e-11& 5.40e-11& 5.30e-11 \\
 3 &  2  &   8.19e-11& 7.88e-11& 7.67e-11& 7.50e-11 \\
 4 &  0  &   1.42e-11& 1.49e-11& 1.43e-11& 1.30e-11 \\
 4 &  1  &   3.55e-11& 3.74e-11& 3.73e-11& 3.49e-11 \\
 4 &  2  &   6.01e-11& 6.22e-11& 6.28e-11& 6.10e-11 \\
 4 &  3  &   5.78e-11& 6.66e-11& 7.01e-11& 6.98e-11 \\
 5 &  0  &   9.94e-12& 9.69e-12& 8.43e-12& 7.43e-12 \\
 5 &  1  &   2.82e-11& 2.95e-11& 2.77e-11& 2.54e-11 \\
 5 &  2  &   4.17e-11& 4.30e-11& 4.13e-11& 3.84e-11 \\
 5 &  3  &   5.98e-11& 6.46e-11& 6.36e-11& 6.06e-11 \\
 5 &  4  &   5.59e-11& 6.40e-11& 6.74e-11& 6.64e-11 \\
\hline
\end{tabular}
\end{table*}

\section{Benchmarking the SUBLIME CO model}
\label{sec:bench}

\begin{figure}
\centering
\includegraphics[width=0.6\columnwidth]{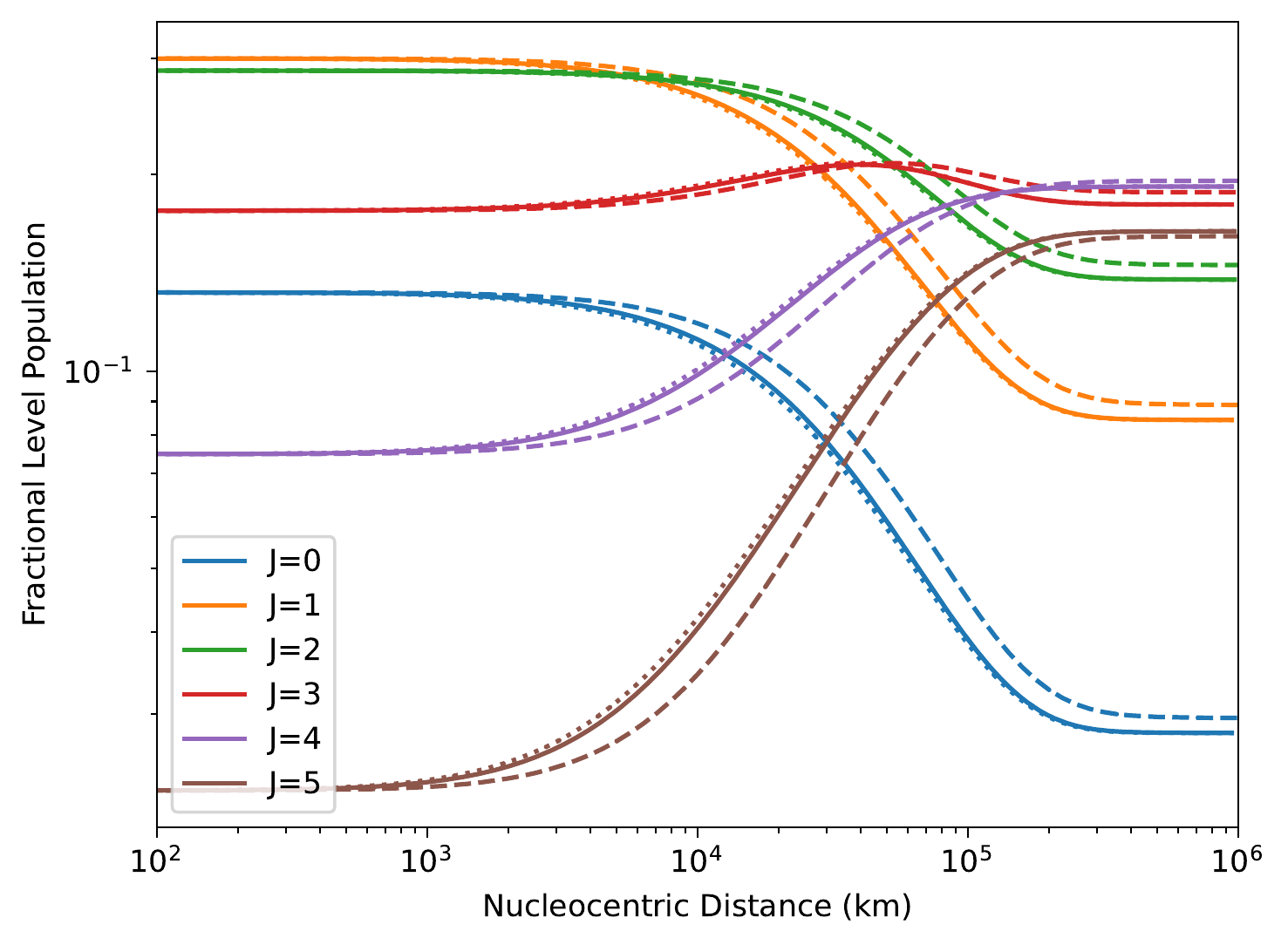} 
\caption{Fractional CO rotational energy level populations for $J=0$--5 as a function of radius, based on a non-LTE, spherically symmetric coma model with $Q({\rm CO})=5\times10^{28}$~s$^{-1}$, $T_{kin}=20$~K, $v_{out}=0.5$~km\,s$^{-1}$ and $r_H=2.8$~au. Solid curves are using the time-dependent version of SUBLIME, with CO--CO collision rates from Section \ref{sec:rates}; dotted curves are assuming the CO--CO collision rates are the same as the CO--H$_2$ rates from \citet{yan10}; dashed curves are using the model of \citet{biv18}.  \label{fig:bench}}
\end{figure}

To confirm the accuracy of our coma radiative transfer and excitation model, we compared results with the similar, well-tested model of N. Biver (private communication). Their model has been used to analyze mm/sub-mm rotational spectra of numerous comets over the last few decades \citep[\emph{e.g.}][]{biv99,boc12,biv16,biv19}, and is based on the formalism presented by \citet{cro87} and \citet{boc87}, described in more detail by \citet{biv97} and \citet{boc04}. Figure \ref{fig:bench} shows a comparison between the computed CO energy level populations from their model and our SUBLIME model, adopting a spherically symmetric (1D) outflow geometry with $Q({\rm CO})=5\times10^{28}$~s$^{-1}$, a constant kinetic temperature $T_{kin}=20$~K and outflow velocity $v_{out}=0.5$~km\,s$^{-1}$, at a heliocentric distance $r_H=2.8$~au. The results of the Biver et al. model are shown with dashed lines, whereas our model is shown with solid lines.

The model populations are in very close agreement considering the complexity of the calculation, and the different assumptions regarding the relevant molecular parameters. Outside of the collisionally dominated (LTE) zone, the models diverge slightly, primarily as a result of the different treatments of CO--CO collision rates --- we are using quantum-mechanically-derived state-specific rate coefficients, whereas their model uses thermalizing rate coefficients based on an (assumed) uniform collisional cross section. Slight differences are also evident at fluorescence equilibrium (largest nucleocentric distances, where the populations reach a steady state with respect to the solar radiation field), presumably due to small differences in the rovibrational Einstein $A$ coefficients used to derive the pumping rates (our model uses the latest HITRAN data).

As a test of the SUBLIME raytracing algorithm, we compared integrated line fluxes from our output model spectral images for two different CO lines, convolved to the JCMT spatial resolution. For the $J=2-1$ line, we obtained $\int{T_Rdv}=0.37$~K\,km\,s$^{-1}$, compared with 0.36~K\,km\,s$^{-1}$ from Biver et al's model, and for $J=3-2$, we have $\int{T_Rdv}=0.64$~K\,km\,s$^{-1}$, compared with 0.63~K\,km\,s$^{-1}$, again, demonstrating very good agreement between our models, at a level much less than the observational uncertainties.

Figure \ref{fig:bench} also shows (with a dotted line style) the level populations derived from the same SUBLIME model, but using CO--H$_2$ rates from \citet{yan10} to approximate the CO--CO collision rates, instead of our new state-specific CO--CO rates calculated in Section \ref{sec:rates} (for $J<6$). At 20~K, the new rate coefficients differ by up to a factor of 7.1 from those of CO--H$_2$, with a mean ratio between the new and H$_2$-derived rates of 2.3. The resulting discrepancy in the final results between coma models using the new rate coefficients, as opposed to adopting CO--H$_2$ rates, however, is relatively small.

\section{Azimuthally averaged JCMT HARP spectra}
\label{sec:azavspec}

\begin{figure}
\centering
\includegraphics[width=0.5\columnwidth]{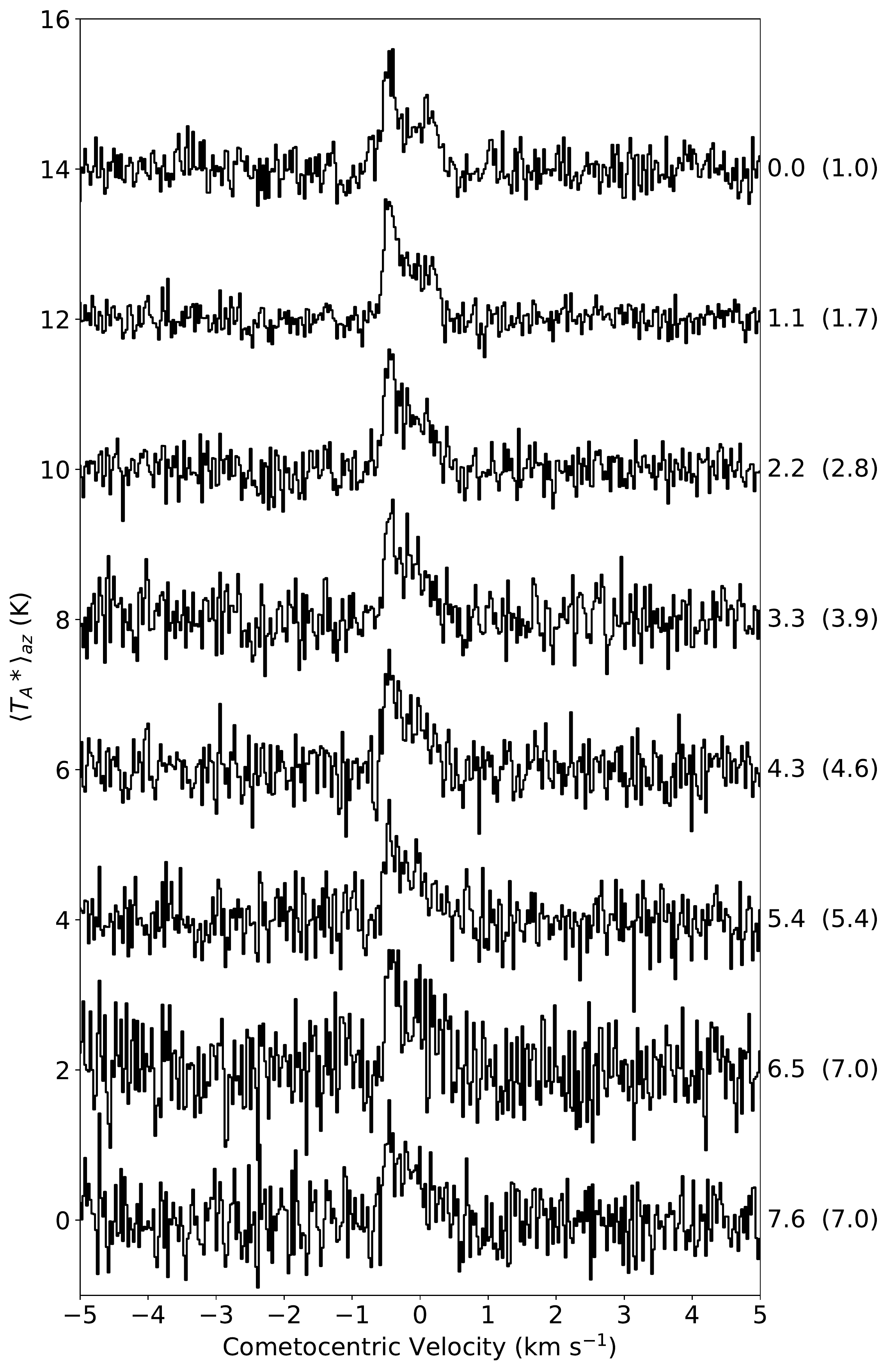} 
\caption{Azimuthally averaged CO $3-2$ spectra, based on JCMT HARP jiggle maps (combined data from 2018-01-14 and 2018-01-15). These are the mean spectra within successive ring-shaped annuli (1 pixel thick), centered on the nucleus, and show the spectral profile as a function of radial offset from the comet (from top to bottom). The numbers to the right of each spectrum indicate the radius of each annulus in units of $10^4$~km. The spectra have each been scaled for display purposes (normalized to the same peak value as the central, 0.0~km spectrum), with scale factors denoted by the numbers in parentheses. \label{fig:azavspec}}
\end{figure}

Azimuthal averages (about the central pixel) of the JCMT HARP CO $3-2$ spectral-spatial data cube are shown in Figure \ref{fig:azavspec}. These are based on the average of the HARP jiggle map observations from 2018-01-14 and 2018-01-15. Each spectrum has been scaled (normalized to the same peak value as the spectrum from the central pixel), to cancel out the rapid decay in the overall line intensity with radius due to the falling coma density. Within the noise, there is no obvious evolution in the spectral line profile with distance from the comet.

\bibliographystyle{aasjournal}
\bibliography{refs.bib}

\end{document}